\documentclass[
reprint,
 amsmath,amssymb,
 aps,
floatfix,
]{revtex4-2}

\usepackage{blindtext}

\usepackage{adjustbox}
\usepackage{subfigure}
\usepackage{amssymb}
\usepackage{upgreek}
\usepackage{verbatim}
\usepackage[separate-uncertainty=true]{siunitx}

\usepackage{graphicx}
\usepackage{dcolumn}
\usepackage{bm}
\usepackage{xcolor}

\begin{document}

\preprint{APS}
\title{Time-resolved optical shadowgraphy of solid hydrogen jets as a testbed to benchmark particle-in-cell simulations}

\author{Long Yang\textsuperscript{1,2}}
\email{yanglong@hzdr.de}
\author{Lingen Huang\textsuperscript{1}}
\author {Stefan Assenbaum\textsuperscript{1,2}}
\author {Thomas E Cowan\textsuperscript{1,2}}
\author {Ilja Goethel\textsuperscript{1,2}}
\author{Sebastian G\"ode\textsuperscript{3}}
\author {Thomas Kluge\textsuperscript{1}}
\author {Martin Rehwald\textsuperscript{1}} 
\author {Xiayun Pan\textsuperscript{1,2}} 
\author {Ulrich Schramm\textsuperscript{1,2}}
\author {Jan Vorberger\textsuperscript{1}}
\author {Karl Zeil\textsuperscript{1}}
\author{Tim Ziegler\textsuperscript{1,2}}
\author{Constantin Bernert\textsuperscript{1,2}}
\email{c.bernert@hzdr.de}

\affiliation{(1) Helmholtz-Zentrum Dresden - Rossendorf, 01328 Dresden, Germany}
\affiliation{(2) Technische Universit\"at Dresden, 01062 Dresden, Germany}
\affiliation{(3) European XFEL GmbH, Schenefeld 22869, Germany}

\date{\today}

\begin{abstract}
Particle-in-cell (PIC) simulations are a superior tool to model kinetics-dominated plasmas in relativistic and ultrarelativistic laser-solid interactions (dimensionless vectorpotential $a_0 > 1$). The transition from relativistic to subrelativistic laser intensities ($a_0 \lesssim 1$), where correlated and collisional plasma physics become relevant, is reaching the limits of available modeling capabilities.  This calls for theoretical and experimental benchmarks and the establishment of standardized testbeds. In this work, we develop  such a suitable testbed to experimentally benchmark PIC simulations using a laser-irradiated micron-sized cryogenic hydrogen-jet target. Time-resolved optical shadowgraphy  of the expanding plasma density, complemented by hydrodynamics  and ray-tracing simulations, is used to determine the bulk-electron temperature evolution after laser irradiation. As a showcase, a study of isochoric heating of solid hydrogen induced by laser pulses with a dimensionless vectorpotential of $a_0 \approx 1$ is presented. The comparison of the bulk-electron temperature of the experiment  with systematic scans of PIC simulations demostrates that, due to an interplay of vacuum heating and resonance heating of electrons, the initial surface-density gradient of the target is decisive to reach quantitative agreement at \SI{1}{\ps} after the interaction. The showcase demostrates the readiness of the testbed for controlled parameter scans at all laser intensities of $a_0 \lesssim 1$.

\end{abstract}

\maketitle

\section{Introduction}
\label{sec:intro}
High-impact applications of high-intensity laser-solid interactions such as fast ignition in inertial-confinement fusion \cite{atzeni2004physics, roth2008review, fernandez2014fast}, laser-driven ion acceleration \cite{daido2012review, macchi2013ion}, the investigation of warm-dense matter \cite{faustlin2010observation,zastrau2014resolving,ren2020observation,chen2014proton,malko2022proton} or secondary-source development \cite{gauthier2017high,liao2019review,rosmej2019interaction,consoli2020laser,treffert2021towards,gunther2022forward} have developed into independent research fields over the last years. Gaining insight into the microscopic interaction picture is the domain of numerical modeling through simulations. In practice, the simulations are most often used retrospectively to guide the analysis and interpretation of exploratory experiments. Prediction making and the development of long-term strategies remains challenging. In translational research in radiation oncology \cite{Baumann2001,Aymar2020,Cirrone2020,Chaudhary2021,kroll2022tumour}, as one prominent example, the  extrapolation of realized proof-of-concept studies towards future high-energy laser-driven proton accelerators \cite{Schwoerer2006,Esirkepov2006} has been delayed \cite{schreiber2016invited,albert20212020}. The technical realization of optimal laser-target interaction conditions and the further development of available simulation tools to correctly capture all occurring physical processes during the transition of the target from an initially solid state to an ultrarelativistic-plasma state represent the key aspects. \\
State-of-the-art modeling of ultrarelativistic laser-solid interactions with Petawatt-class lasers (peak intensity $\gtrsim 10^{21} \SI{}{\W / \cm^2}$) is currently based on a staged approach of numerical simulations \cite{Schollmeier2015, Nishiuchi2020, dover2023enhanced} that includes target preexpansion by the leading edge of the high-power laser \cite{Danson2019}. For  intensities in the vicinity of the laser peak, i.e., dimensionless vectorpotentials $a_0 \gtrsim 1$, particle-in-cell (PIC) simulations are optimized to compute the kinetic regime of particle motion and thermal nonequilibrium. Adequate initial conditions of the PIC simulations are derived by radiation-hydrodynamics simulations, which compute the interaction in the subrelativistic intensity regime ($a_0 \ll 1$) during the leading edge. \\
However, the transition from  relativistic to  subrelativistic laser intensities, i.e., dimensionless vectorpotentials $a_0 \lesssim 1 $ ($10^{16}$ to $10^{18} \, \SI{}{\W / \cm^2}$ for \SI{800}{\nm} light), is currently reaching the limits of available modeling capabilities. The two most obvious approaches to cover this laser-intensity regime are being pursued; the extension of hydrodynamics-simulation tools \cite{eidmann2000hydrodynamic} and  the inclusion of 
 correlated and collisional plasma physics into PIC-simulation frameworks \cite{arber2015, albert20212020}. These developments call for standardized theoretical and experimental benchmarks  under unified interaction scenarios \cite{Smith2021}. As microscopic parameters are usually not directly accessible from laser-plasma interactions, such benchmarks would ideally require macroscopic observables that allow for an unambiguous allocation of specific interaction conditions. \\
 
In this work, we present a testbed to experimentally benchmark PIC simulations based on a laser-irradiated micron-sized cryogenic hydrogen-jet target \cite{kim2016development,Curry2020,Rehwald2023}. The corresponding PIC simulations of the laser-target  interaction benefit from the comparably low target density \cite{gode2017relativistic,obst2017efficient,obst2018all,polz2019efficient, Rehwald2023a}, the single-species composition, the negligible amount of Bremsstrahlung radiation (atomic number $Z=1$)  and simple ionization dynamics \cite{Bernert2023}. Furthermore, the plasma composition of only protons and electrons enhances comparability to analytic calculations. The temporal evolution of the laser-heated plasma density is visualized by time-resolved off-harmonic optical probing \cite{Bernert2022} via two spectrally separated, ultrashort laser beams. A fitting approach by hydrodynamics and ray-tracing simulations completes the testbed. It enables the determination of the bulk-electron temperature evolution after the laser energy was absorbed by the target. \\
The capabilities of our testbed are showcased in an experiment studying isochoric heating of solid hydrogen when irradiated by laser pulses with \SI {37}{\fs} duration and $a_0 \approx 1$. The focal spot of the laser is kept larger than the spatial dimensions of the target to guarantee a homogeneous interaction.  Isochoric heating of solids by high-intensity lasers is a widely-used approach to generate warm-dense and hot-dense matter \cite{saemann1999isochoric,perez2010enhanced,martynenko2021role}. Therefore, the process is particularly well suited to benchmark PIC-simulation results.  During  isochoric heating, the ultra-short  laser pulse accelerates  electrons on the laser-plasma interface up to relativistic velocities \cite{beg1997study,wilks2001energetic,kluge2011electron,kluge2018simple,singh2022vacuum}. Subsequently, the electrons traverse  the target and generate a thermalized bulk-electron and bulk-ion temperature, mainly by resistive heating, drag heating or diffusion heating \cite{huang2013ion,kemp2006collisional,huang2016dynamics,chrisman2008intensity}. Finally, the plasma undergoes an adiabatic expansion into vacuum and the electron and ion temperatures equilibrate \cite{fletcher2022electron}. For lasers with a dimensionless vectorpotential $a_0 \lesssim 1$, a transition from vacuum heating to resonance heating of electrons exists, which depends on the surface-density scalelength of the target \cite{gibbon1992collisionless,brunel1987not,gibbon2005short,kruer2003physics,azamoum2018impact}. Furthermore, for $a \lesssim 1$, the collisionality of particles increases in relevance. \\
 The  bulk-electron temperature after isochoric heating represents a specifically feasible endpoint, i.e., a macroscopic observable, to benchmark the entire modeling chain of nonthermal equilibrium  that can be computed by PIC simulations, including laser acceleration of electrons and their thermalization to a single temperature within the target bulk. Experimentally, the determination of the achieved bulk-electron temperature most often relies on spatially and temporally integrated x-ray spectroscopy  \cite{huang2013ion,de1981soft,griem2005principles,nilson2009bulk,renner2019challenges,martynenko2020effect}. Time-resolved optical shadowgraphy of the induced expansion  provides a measure of the bulk-electron temperature subsequent to isochoric heating \cite{bang2015visualization,bang2016linear}. \\

After introducing the testbed and the method for determining the bulk-electron temperature from the experimental data, the results are compared to systematic scans of PIC simulations. We find that, due to the interplay of vacuum heating and resonance heating of electrons, the bulk-electron temperature is highly sensitive to the surface-density gradient of the target. By demonstrating the capability of the testbed to quantitatively benchmark PIC simulations, it establishes the basis towards controlled parameter scans at laser intensities well below $a_0 = 1$. This transition regime from solid state to ultrarelativistic plasmas is highly relevant to all high-intensity laser-solid interaction and their applications.

\section{Testbed platform}
 This section presents the testbed to experimentally benchmark PIC simulations of high-intensity laser-solid interactions via the endpoint of the bulk-electron temperature after isochoric heating. The endpoint includes a chain of physical processes, i.e., target ionization, the laser-acceleration of electrons as well as their thermalization within the initially cold target bulk to a single Maxwellian electron temperature.  The bulk-electron temperature well above the Fermi energy induces an adiabatic plasma expansion into vacuum, which is visualized by time-resolved optical diagnostics. Section \ref{sec:experiment} presents the time-delay scan of two-color shadowgraphy in the tens-of-picosecond timeframe together with the utilized experimental setup. The determination of the corresponding electron-temperature evolution is explained in section \ref{sec:temperature_fit}. Section \ref{subsec:hydro} assumes a homogeneous electron temperature throughout the central plane of the target and calculates the electron-density evolution via a hydrodynamics simulation. Section \ref{subsec:ray_tracing} simulates the corresponding shadow diameters versus delay by ray-tracing simulations and  section \ref{subsec:fit_to_experimental_data} presents a $\chi^2$ fit of the simulated to the measured shadow diameters. By this we derive the best-matching hydrodynamics simulation and the corresponding electron-temperature evolution that corresponds to the bulk-electron-temperature evolution after isochoric heating and thermalization. The combination of hydrodynamics simulations, ray-tracing simulations and the $\chi^2$ fit is referred to as \textit{HD-RT fit} in the following.

\label{sec:testbed}
\subsection{Experiment}
\label{sec:experiment}

\begin{figure}[htb] 
\centering {
\includegraphics[width=83mm]{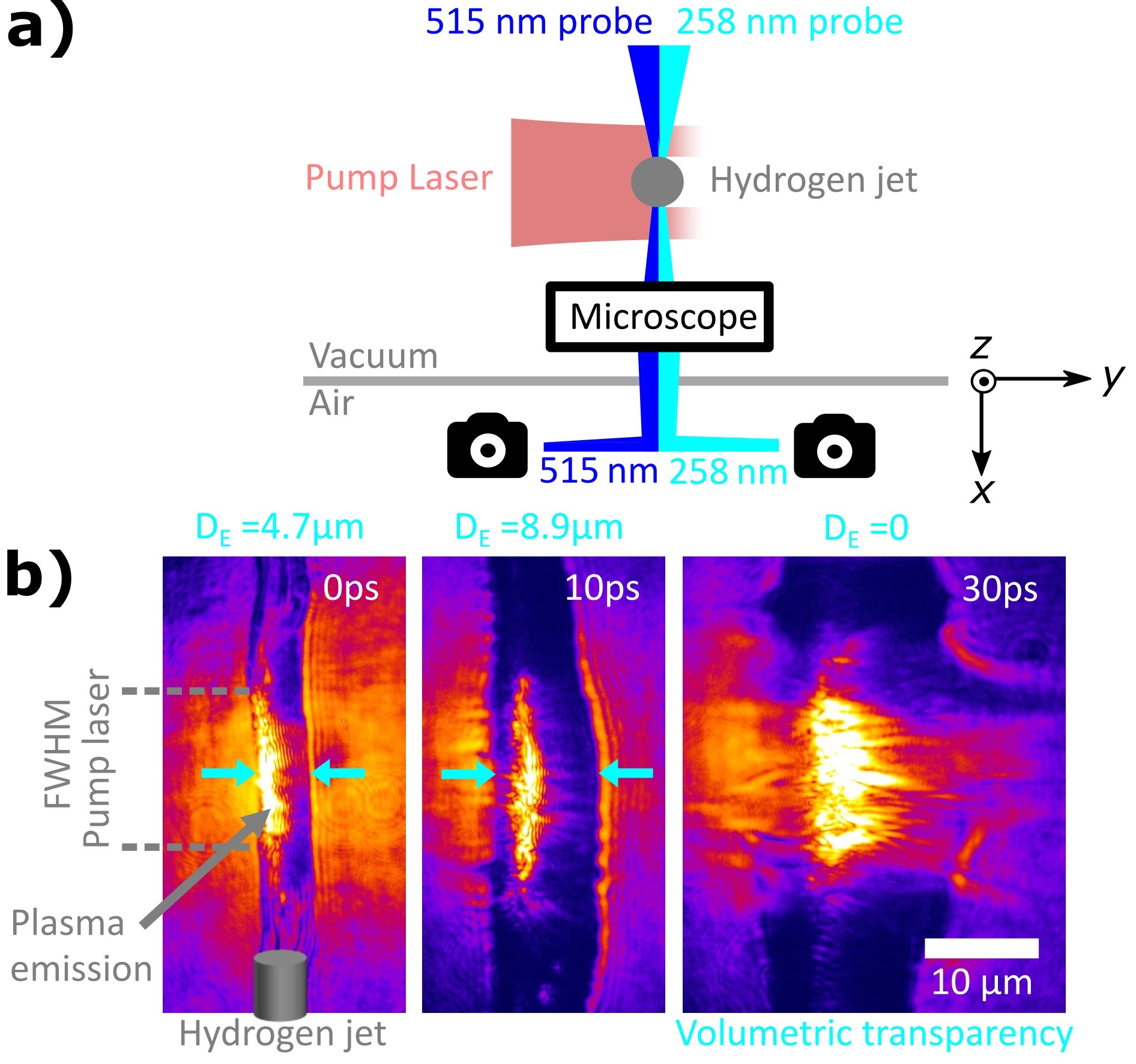}} 
\caption{\textbf{a)} Experimental setup: The \SI{37}{fs}-pump laser with a peak intensity of \SI{1.6E18}{W/\cm^2} interacts with a  cylindrical hydrogen-jet target. Time-resolved high-resolution shadowgraphy utilizes a single microscope and two simultaneous backlighters with a pulse duration of \SI{260}{fs} (\SI{258}{nm} probe) and \SI{160}{fs} (\SI{515}{nm} probe). \textbf{b)} Exemplary shadowgrams of a time-delay scan as recorded by the \SI{258}{nm} probe. The pump-probe delay is given in the upper-right corner of each shadowgram and the shadow diameter $D_\text{E}^{\SI{258}{nm}}$ is measured between the cyan arrows. For \SI{30}{ps} delay, \textit{volumetric transparency} is observed. The spatial extent and fluence of the parasitic plasma emission varies from shot to shot.
The colorscales and spatial scales are consistent between the images.}
\label{Fig.2}
\end{figure}

The experiment is conducted with the Draco-\SI{150}{TW} laser  at the Helmholtz-Zentrum Dresden-Rossendorf \cite{schramm2017first}. The top view of the setup is shown in figure \ref{Fig.2} (a). By placing  a \SI{16}{mm}-wide circular aperture in the collimated beam path, the pump-laser energy is reduced to \SI{160}{mJ}.  Each laser pulse features a  duration of \SI{37}{fs} full width at half maximum (FWHM) and a central wavelength of $\lambda_\text{laser} = \SI{800}{nm}$. It is focused by a f/16 off-axis parabola (OAP) and generates an Airy-pattern focus with a spatial FWHM of \SI{14}{\um} of the central disk and a peak intensity of $I_\text{laser} = \SI{1.6E18}{\W\cm^{-2}}$. \\
A continuous, self-replenishing jet of cryogenic hydrogen is used as a target \cite{kim2016development,Curry2020}. The cross section of the target at the source is defined by a circular aperture with a  diameter of \SI{5\pm1}{\um}.  At the position of the laser-target interaction, the mean diameter is \SI{4.4}{\um} with a standard deviation of \SI{0.2}{\um} (details in Appendix \ref{app:initial_diameter}). The electron density of the fully ionized target is \SI{5.2E22}{cm^{-3}}, which corresponds to \SI{30}{} times the critical plasma density of \SI{800}{nm} light $n_c^{\SI{800}{\nm}}$.\\
The laser-target interaction is investigated by time-resolved off-harmonic optical shadowgraphy  \cite{Bernert2022} at \SI{90}{\degree} angle to the pump-laser propagation direction. Two copropagating backlighter pulses with the wavelengths $\lambda$ of \SI{515}{nm} and \SI{258}{nm} are generated from a synchronized stand-alone laser system \cite{Loeser2021}. The \textit{258 nm probe} and the \textit{515 nm probe} feature a pulse energy of approximately \SI{0.3}{\micro J} and \SI{5}{\micro J}  and a pulse duration of \SI{260}{fs} and \SI{160}{fs}, respectively. The pump-probe delay is variable, the temporal resolution of a time-delay scan is \SI{175}{fs} and the captured data is blurred by the pulse duration \cite{Bernert2022}. Dispersion effects in the beam path \cite{Kaluza2008} cause a fixed delay of \SI{4.6}{\ps} between both probe-laser pulses. All delays are given with respect to the arrival of the pump-laser peak on target at \SI{0}{\ps} delay. To reduce the influence of parasitic plasma emission, both probe beams are focused by a f/1 OAP. \\
A long-working-distance infinite-conjugate microscope objective (designed for laser wavelengths of \SI{266}{nm} and \SI{532}{nm}) is used to image the shadow of the target simultaneously for both wavelength, i.e., at two different ranges of plasma density. The two images are separated by a dichroic mirror and recorded by different cameras, each equipped with the corresponding color filters (more details in Appendix \ref{app:exp_details}). \\

\begin{figure}[htb] 
\centering {
\includegraphics[width=83mm]{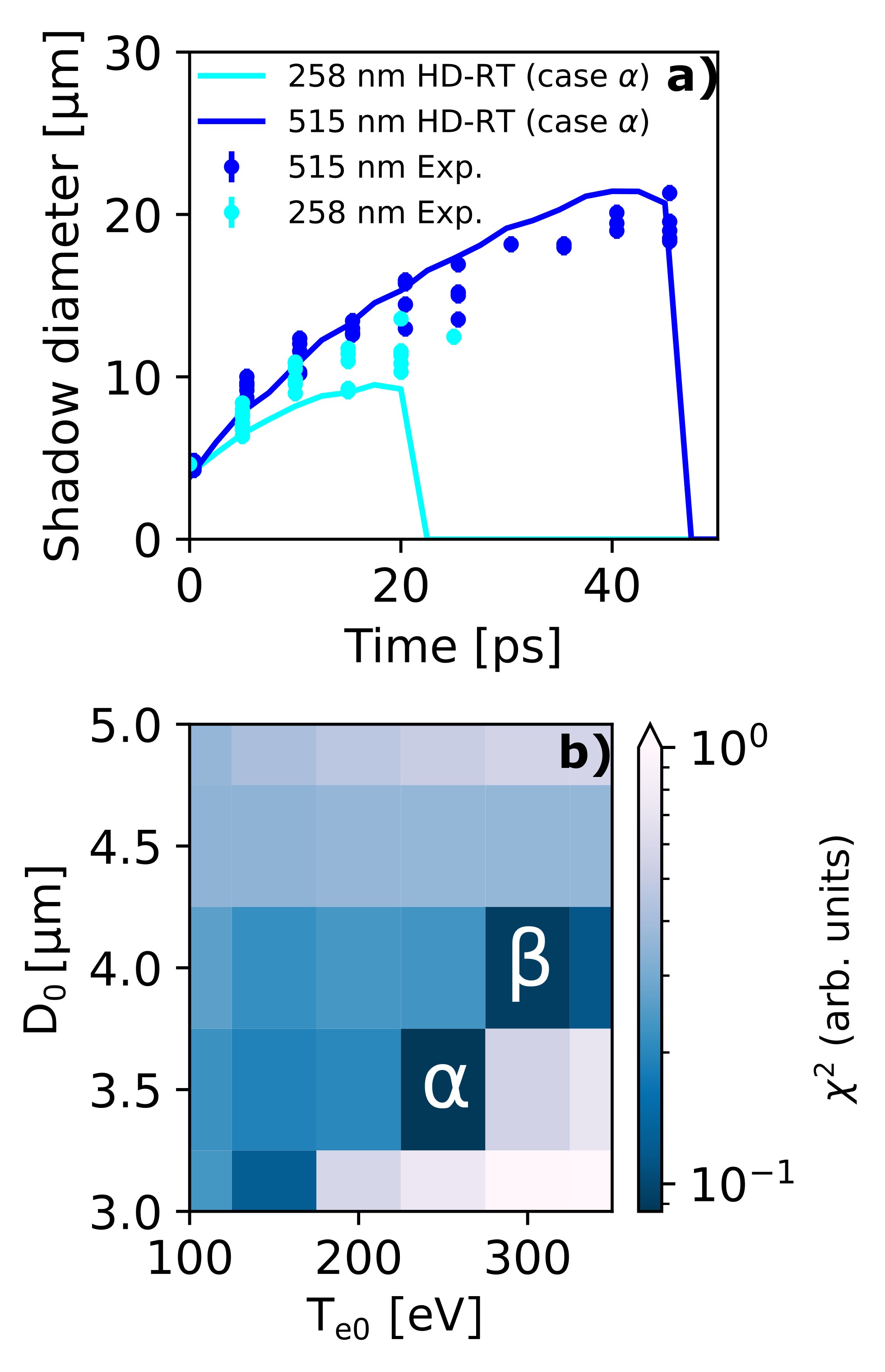}} 
\caption{\textbf{a)} Measured shadow diameter from two-color optical shadowgraphy and  HD-RT fit  to the experimental data (case $\alpha$, see section \ref{subsec:fit_to_experimental_data}). \textbf{b)} The total $\chi ^2$ distribution between the experimental data and all HD-RT simulation data.}
\label{Fig.1}
\end{figure}

A pump-probe time-delay scan in steps of \SI{5}{ps} is performed. The inherent spatial jitter of the target is controlled by a secondary optical-probe axis (not shown here) \cite{obst2017efficient}. Only data from within \SI{\pm2}{\um} of the object plane is presented. Exemplary shadowgrams at \SI{0}{ps}, \SI{10}{ps} and \SI{30}{ps} delay are shown in figure \ref{Fig.2} (b). The images are recorded by the \SI{258}{\nm} probe and show a sideview of the target.  The pump laser impinges from the left and the spatial FWHM of the focal spot is indicated by the dashed lines on the left-hand side. Plasma emission occurs mostly from within the FWHM of the focal spot  and from the position of the unperturbed target-front surface. The shadow of the target features sharp edges.  The shadow diameter $D_\text{E}^{\SI{258}{nm}}$ measures the width of the shadow at the position of the pump-laser peak, as illustrated by the cyan arrows in the shadowgrams at \SI{0}{ps} and \SI{10}{ps} delay. At \SI{30}{ps}, the edges of the shadow are diffuse and prevent the measurement of $D_\text{E}^{\SI{258}{nm}}$. Consequently, $D_\text{E}^{\SI{258}{nm}}$ is set to zero for this delay. The shadowgram shows penetration of probe light through central parts of the target. In the following we refer to this observation as \textit{volumetric transparency}. \\
The whole scan of shadow diameter versus pump-probe delay is shown by the markers in figure \ref{Fig.1} (a). $D_\text{E}^{\SI{258}{nm}}$ and $D_\text{E}^{\SI{515}{nm}}$ both increase with pump-probe delay until volumetric transparency sets in. For the \SI{258}{\nm} probe and the \SI{515}{\nm} probe, this occurs between \SI{20}{} and \SI{25}{\ps} and between \SI{45}{} and \SI{50}{\ps}, respectively. For all delays greater than zero, $D_\text{E}^{\SI{515}{nm}}$ is on average larger than $D_\text{E}^{\SI{258}{nm}}$. This is expected, as the critical plasma density $n_c$ drops with increasing wavelength $\lambda$ according to
\begin{equation}
    n_c^\lambda \propto \frac{1}{\lambda^2} \, .
\end{equation}
Assuming an exponentially decreasing plasma-density surface, the difference between $D_\text{E}^{\SI{258}{nm}}$ and $D_\text{E}^{\SI{515}{nm}}$ enables the calculation of the corresponding scalelength $L_0$. For \SI{0}{\ps} delay, the surface scalelength calulates to values between \SI{0}{} and about \SI{0.42}{\um}. Figure \ref{Fig.1} (a) shows shot-to-shot fluctuations of the shadow diameter. They are caused by inherent shot-to-shot variations of the target and the pump laser. Target variations include local  changes of diameter and aspect ratio on the submicron scale as well as bow-like deformations along the jet axis on the micron scale (details in Appendix \ref{app:initial_diameter}). Furthermore, the rapid freezing of the jet introduces compositional and structural variations of the solid hydrogen \cite{Kuehnel2011}. The main source of fluctuation of the pump-laser intensity is the peak power with a measured standard deviation of \SI{12}{\percent}. The fluctuation of the pump-laser energy is about \SI{1}{\percent} and the pointing jitter is negligible.

\subsection{HD-RT fit}
\label{sec:temperature_fit}

\subsubsection{Hydrodynamics simulation - HD}
\label{subsec:hydro}

\begin{figure*}[htb] 
\centering {
\includegraphics[width=166mm]{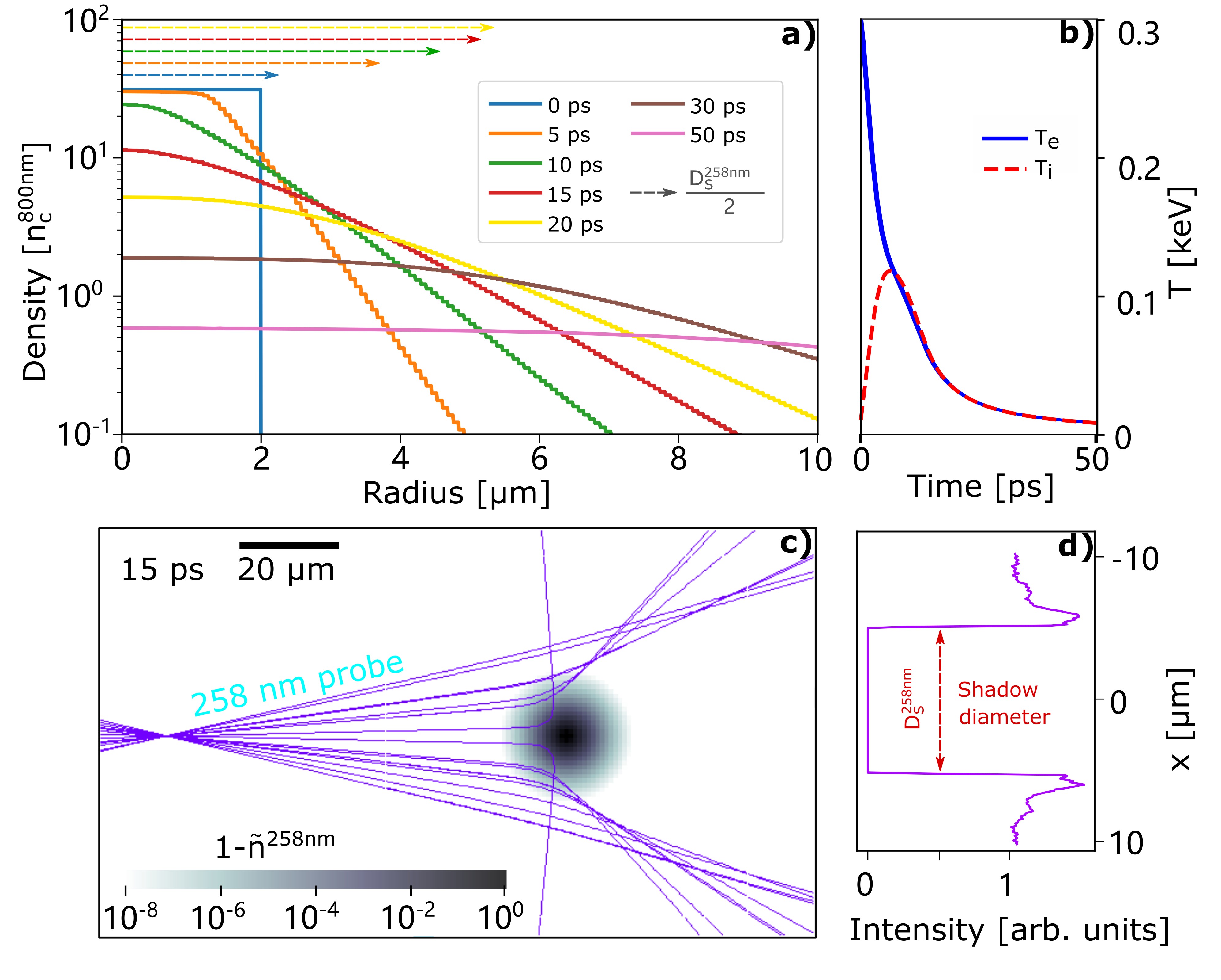}} 
\caption{Results of a hydrodynamics simulation and ray-tracing simulations: \textbf{a)} Radial electron density of the hydrodynamics simulation with an initial setting of $T_{e0}$ = \SI{300}{\eV} and $D_{0}$ = \SI{4}{\um} (case $\beta$ in figure \ref{Fig.1} (b)). The corresponding shadow diameters $D_\text{S}^{\SI{258}{\nm}}$ are derived by  ray-tracing simulations and are given by the dashed arrows on the top. \textbf{b)} Electron and proton temperature of the target bulk at zero radius versus time (case $\beta$ in figure \ref{Fig.1}(b)). \textbf{c)} Top view of the distribution of refractive indices $\tilde{n}^{\SI{258}{\nm}}$ (gray colorscale), which is calculated from the radial electron-density profile at \SI{15}{\ps} delay. The distribution is placed in the object plane of the ray-tracing simulation and refraction of the \SI{258}{\nm}-probe rays is visualized by the purple lines. \textbf{d)} Light distribution in the image plane of the ray-tracing simulation. The shadow diameter $D_\text{S}^{\SI{258}{\nm}}$ is measured at half of the unperturbed background intensity. }
\label{fig:hydro_ray_tracing}
\end{figure*}

We assume that the density evolution of the hydrogen plasma is driven by a two-temperature adiabatic expansion, which can be modeled via  hydrodynamics simulations. In the following, the simulation tool FLASH \cite{fryxell2000flash,dubey2009extensible} is used to perform two-temperature hydrodynamics simulations in two-dimensional radial symmetry with the hydrogen equation of state FPEOS \cite{hu2011first} (details in Appendix \ref{app:hydrogen_equation_of_state}). Initially, there are three free parameters of the target model: the ion temperature $T_{i0}$, the electron temperature $T_{e0}$ and the initial plasma diameter $D_0$. Because the evolution of the shadow diameter is sensitive to the initial plasma diameter $D_0$, which is subject to experimental fluctuations, we handle $D_0$ as a free parameter within the range of the experimental uncertainties. Compared to the effect of $T_{e0}$, the initial ion temperature $T_{i0} \ll T_{e0}$ has little effect on the expansion process. As derived by PIC simulations in section \ref{sec:PICsimulation}, $T_{i0}$ can be approximated by \SI{10}{\eV}. \\
Hydrodynamics simulations with different initial electron temperatures $T_{e0}$ and different initial plasma diameters $D_0$ are conducted. Each simulation box has a length of \SI{50}{\um} and a vacuum density of \SI{1e-8}{} times the target density. An exemplary evolution of the electron density is shown in figure \ref{fig:hydro_ray_tracing} (a) and the corresponding evolution of ion and electron temperature is shown in figure \ref{fig:hydro_ray_tracing} (b). 

\subsubsection{Ray-tracing simulation - RT}
\label{subsec:ray_tracing}

To compare the evolution of the electron density in the hydrodynamics simulation with the experimentally measured shadowgraphy data, the shadow formation of each probe beam needs to be modeled. For optical shadowgraphy of solid-density plasmas, shadow formation is governed by refraction on the density gradients of the under-critical regions of the plasma \cite{Bernert2022}. Each simulated density profile is transformed into a two-dimensional distribution of refractive indices $\tilde{n}$ from the local electron density $n_e$ via the dispersion relation \cite{Macchi2013} 
\begin{equation} 
\tilde{n} = \sqrt{1-\frac{n_e}{n_c}} \,.
\end{equation}
 The critical density $n_c$ depends on the  wavelength and $\tilde{n}^{\SI{258}{\nm}}$ and $\tilde{n}^{\SI{515}{\nm}}$ are calculated separately.  \\
The experimental imaging setup and the beam path of each probe are reproduced in a virtual optical setup with the software Zemax \footnote{Zemax 13 Release 2 SP6 Professional (64-bit)}.  The calculated spatial distribution of refractive indices  $\tilde{n}^{\SI{258}{\nm}}$ or $\tilde{n}^{\SI{515}{\nm}}$ is inserted into the object plane of the corresponding setup, which is exemplary shown in figure \ref{fig:hydro_ray_tracing} (c). The purple lines illustrate the refraction of the \SI{258}{\nm}-probe rays in the gradients of the refractive-index distribution. The light distribution in the image plane is calculated by Zemax and presented in figure \ref{fig:hydro_ray_tracing} (d). The graph shows the formation of a shadow with sharp edges, similar to the experiment. Refraction leads to an increased intensity level at the rim of the shadow edges.  The shadow diameter $D_\text{S}^{\SI{258}{\nm}}$ is derived at half of the unperturbed background intensity (\mbox{\SI{0.5}{} arb. units}). By utilizing the virtual setup of the \SI{515}{\nm} probe, $D_\text{S}^{\SI{515}{\nm}}$ is calculated accordingly. \\
The simulated evolution of $D_\text{S}^{\SI{258}{\nm}}$ (cyan) and $D_\text{S}^{\SI{515}{\nm}}$ (blue) of a hydrodynamics simulation with the initial setting $T_{e0}$ = 250 eV and $D_0$ = \SI{3.5}{\um} (case $\alpha$) is shown as solid lines in figure \ref{Fig.1} (a). Like in the experiment, $D_\text{S}$ is set to zero at delays for which no sharp shadow edge is derived and volumetric transparency is observed instead. For the \SI{258}{\nm} probe and the \SI{515}{\nm} probe of case $\alpha$, this occurs at \SI{22.5}{\ps} and \SI{47.5}{\ps} delay, respectively. 

\subsubsection{$\chi^2$ fit}
\label{subsec:fit_to_experimental_data}

To find the best-matching hydrodynamics simulation to the experimental data, the initial electron temperature $T_{e0}$ is varied from \SI{100}{} to \SI{350}{\eV} in steps of \SI{50}{\eV}. Furthermore, the initial plasma diameter $D_0$ is scanned from \SI{3}{} to \SI{5}{\um} in steps of \SI{0.5}{\um}. The variance $\chi ^2$ of the difference between the experimental  data $D_\text{E}(t,\lambda)$ and the simulation data $D_\text{S}(t,\lambda,T_{e0},D_0)$ is  
\begin{equation}
\chi ^{2}(T_{e0},D_0)= \sum_{t,\lambda}\left ( D_\text{E}(t,\lambda)-D_\text{S}(t,\lambda,T_{e0},D_0) \right )^2 \, . 
\end{equation}
Here, $t$ is the pump-probe delay and $ \lambda $ is the wavelength of the probe. The resulting $\chi^2$ map is shown in figure \ref{Fig.1} (b). Minimum values of $\chi^2$ are reached for case $\alpha$ ($T_{e0}$ = 250 eV, $D_0$ = \SI{3.5}{\um}) and case $\beta$ ($T_{e0}$ = 300 eV, $D_0$ = \SI{4}{\um}). Case $\beta$ has better agreement to the \SI{258}{\nm}-probing data and case $\alpha$ has better agreement to the \SI{515}{\nm}-probing data. The two cases give a lower and upper limit of the best-fitting $T_{e0}$ and $D_0$.  A discussion of the method of the HD-RT fit is given in Appendix \ref{app:discussion_HD_RT_fit}. \\

In summary, the HD-RT fit allows to fit a heuristic electron-temperature evolution subsequent to isochoric heating and thermalization of the bulk electrons, which constitutes the endpoint for the comparison to PIC simulations in the presented testbed. For the here presented experimental data, the HD-RT fit  yields a heuristic initial bulk-electron temperature $T_{e0}$ between \SI{250}{} and \SI{300}{\eV} at \SI{0}{\ps} delay.

\section{PIC simulation}
\label{sec:PICsimulation}

 \begin{figure*}[htb] 
\centering {
\includegraphics[width=166mm]{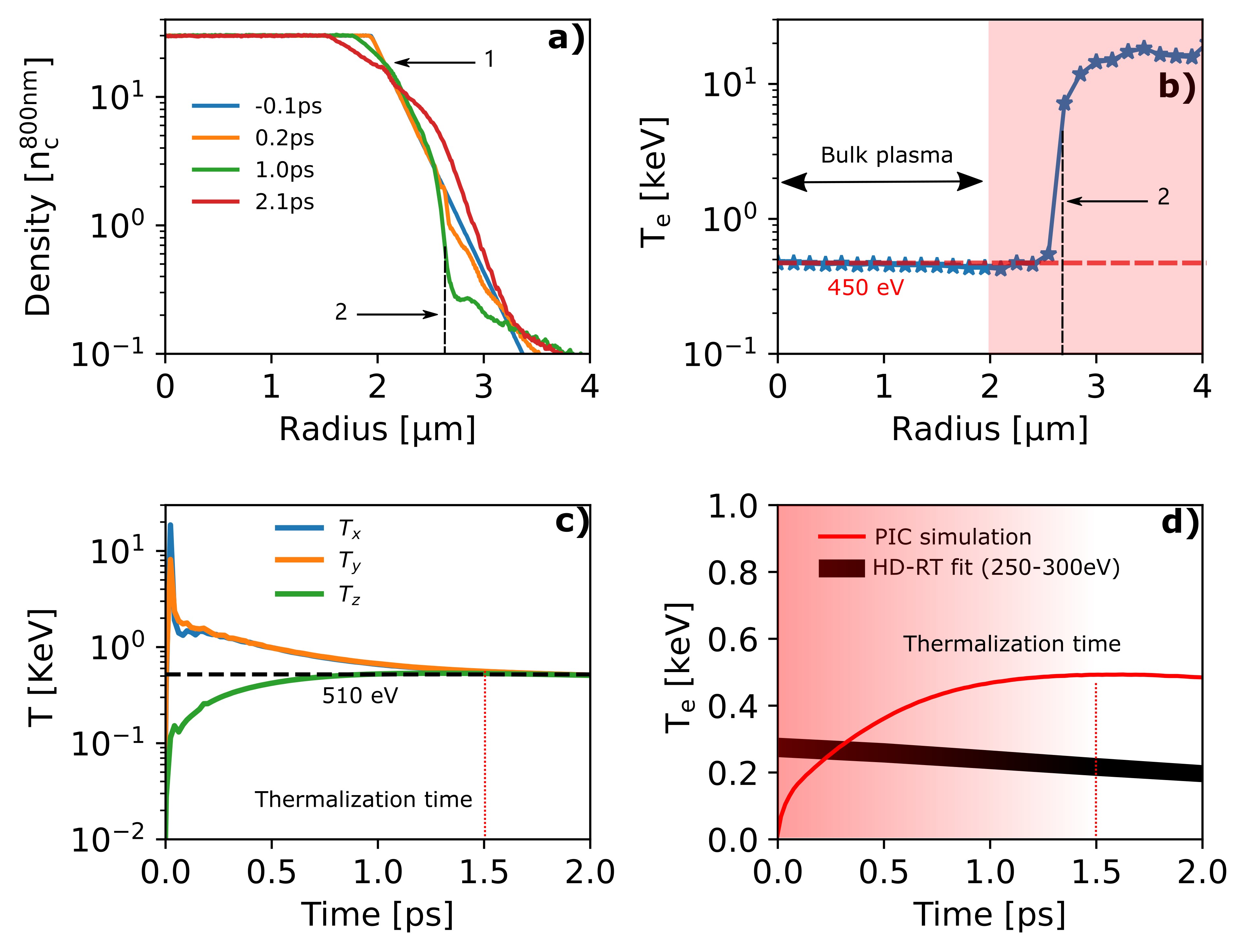}} 
\caption{\textbf{a)} Spatial distribution of the electron density in the PIC simulation. \textbf{b)} Spatial distribution of the electron temperature $T_e$ at \SI{1}{\ps} delay. \textbf{c)} Temporal evolution of the equivalent temperatures $T_n$ (equation \ref{eq:average_kin_energy}). \textit{Thermalization} is reached at about \SI{1.5}{\ps}. \textbf{d)} Comparison of the PIC simulation and the HD-RT fit to the experiment.}
\label{Fig.3}
\end{figure*}

This section exemplary demonstrates a comparison of the results of the testbed with a PIC simulation of the high-intensity laser-solid interaction.  The comparable endpoint is the temporal evolution of the bulk-electron temperature. The presentation of the results of the PIC simulation details the processes of isochoric heating and thermalization of the bulk electrons (not captured by the hydrodynamics simulation of the HD-RT fit). Both finally lead to the evolution of the bulk-electron temperature after thermalization, which can be compared to the hydrodynamics simulation of the HD-RT fit. \\
The simulation is performed with the 2D3V-PIC-simulation tool PIConGPU \cite{bussmann2013radiative}. The target is modeled by a fully ionized spherical hydrogen column. The density distribution follows the model 
\begin{equation}
n(r) =n_0 \,( \Theta(r_c-r)+ \Theta(r-r_c) \, e^{-(r-r_c)/L_0})
\end{equation}
with $n_0=30 \, n_c^{\SI{800}{\nm}}$, a surface scalelength $L_0 = \SI{0.25}{\um}$ (mean of the experimental uncertainty), an unperturbed target diameter of $D_0 =\SI{4.4}{\um}$ (see Appendix \ref{app:initial_diameter}) and the reduced radius 
\begin{equation}
r_c=\sqrt{\frac{D_0^2}{4}-L_0^2}-L_0
\end{equation}
to fulfill mass conservation. The initialized total length of the surface plasma corresponds to ten times $L_0$.   The laser is incident along the $y$ direction and polarized in $x$ direction. It features a Gaussian shape in space and time, a  pulse duration of \SI{37}{\fs} FWHM and a laser-spot size of \SI{14}{\um} FWHM. The simulation box is $\SI{32}{} \times \SI{32}{\um^2}$ with absorbing boundary conditions and the simulation uses a cell length of $\SI{0.8}{\um}/96$ and \SI{240}{} macro particles per cell. The simulation includes a relativistic-binary-collision mechanism. \\
 Figure \ref{Fig.3} (a) shows the simulated electron-density profile for different delays. \SI{0}{\ps} delay corresponds to the arrival of the pump-laser peak on the front side of the target. The comparison of the blue and the orange line  shows that the electron density at \SI{0.2}{\ps} delay does not significantly differ from the initial state at \SI{-0.1}{\ps}. The expansion of hot electrons into the surrounding vacuum occurs at densities below about $ 0.1 \, n_c^{\SI{800}{\nm}}$. Comparing the electron-density profiles at \SI{0.2}{\ps}, \SI{1.0}{\ps} and \SI{2.1}{\ps}, two characteristics are identified. The density region denoted by ``1'' shows a temporally increasing exponential scalelength of the plasma density. It results from adiabatic expansion, which is driven by the electron temperature $T_e$ inside the target bulk (refer to ``Bulk plasma'' in fig. \ref{Fig.3} (b)).  For the electron-density profile at \SI{1.0}{\ps} delay, however, the exponential scalelength is overlayed by a transiently occurring  step of the density profile that is denoted by ``2''. It is caused by the thermal pressure of the coronal plasma that features a much higher temperature than the bulk plasma. The spatial distribution of the electron temperature $T_e$ at \SI{1.0}{\ps} delay is displayed in figure \ref{Fig.3} (b). The electrons inside the bulk plasma feature a spatially constant temperature of \SI{450}{\eV} while the coronal plasma features temperatures above \SI{10}{\keV}. For comparison of figures \ref{Fig.3} (a)  and (b), the vertical dashed line indicates the radius of the strong increase of $T_e$, which coincides with the step of the electron-density profile denoted by ``2''.  As the hotter coronal plasma expands faster than the bulk plasma, the influence of the coronal plasma on the plasma densities in the range above $0.1 \, n_c^{\SI{800}{\nm}}$ decreases with delay and adiabatic plasma expansion most likely dominates for longer delays (compare the evolution of regions ``1'' and ``2'' of the density profiles at \SI{1.0}{\ps} and \SI{2.1}{\ps} delay). Note that the hydrodynamics simulations of the HD-RT fit considers adiabatic plasma expansion only and agreement between the density profiles of the PIC simulation and the HD-RT fit to the experiment is expected only at delays for which region ``1'' dominates the expansion process. It follows that, in the discussed case, the evolution of the electron temperature $T_e$ in the bulk plasma is a better suited endpoint of the comparison than a direct comparison of the density profiles. \\
The temporal evolution of the bulk-electron temperature $T_e$ and the \textit{equivalent temperatures} to the  average kinetic energy $T_x$, $T_y$ and $T_z$ are displayed in figures \ref{Fig.3} (d) and (c).  $T_n$ with $n = x$, $y$ or $z$ is calculated from the average kinetic energy 
\begin{equation}
\overline{E}_{kn}= \frac{1}{2} \, k_B \, T_n 
\label{eq:average_kin_energy}
\end{equation}
 of the electrons within the bulk plasma (circle with a radius of \SI{2}{\um}). As the laser is polarized into the $x$ direction and propagates along the $y$ direction, $T_x$ and $T_y$ include the contribution of non-thermal laser-accelerated electrons. In contrast, $T_z$ is generated by collision of particles only. For 2D3V-PIC simulations, the bulk-electron temperature $T_e$ is close to the temperature component $T_z$.
 Throughout this work,  the quantity $T_e$ of the PIC simulations is derived by a Maxwellian fit to the electron-velocity distribution of the bulk plasma. \\
 The maximum of $T_x$ and $T_y$ of \SI{20}{keV} is reached at \SI{67}{\fs}. Together with the approximately constant electron-density profile between \SI{-0.1}{} and \SI{0.2}{\ps}, this indicates that the heating of the plasma bulk happens isochorically. Subsequently, as the plasma starts to expand, the temperatures $T_x$ and $T_y$ decrease while $T_z$ increases. The term \textit{thermalization} refers to the circumstance that all electrons  get the same Maxwellian temperature distribution into all spatial dimensions via collisions. As figure \ref{Fig.3} (c) shows, $T_x$, $T_y$ and $T_z$ equal each other after about \SI{1.5}{ps}, i.e., the plasma thermalized within about \SI{1.4}{ps} after the termination of heating.  From the analytic electron-electron collision rate of hot electrons and assuming a plasma with an electron density of $30 \, n_c^{\SI{800}{nm}}$ and a temperature of \SI{20}{\keV} we find a thermalization time \cite{richardson20192019}
\begin{equation}
\tau= \frac{1}{\nu_{ee}} = \SI{1.0}{ps} \, ,
\end{equation}
 which is in close agreement to the PIC simulation. We emphasize that the testbed utilizes a pure hydrogen target, which allows for the comparison to analytical calculations without further approximations. \\ 
Figure \ref{Fig.3} (d) compares the evolution of the bulk-electron temperature $T_e$ of the PIC simulation (red line) and the electron-temperature evolution of the HD-RT fit to the experiment (black line). The lower and upper limit of the black line correspond the cases $\alpha$ and $\beta$. Up to \SI{1.5}{\ps} the PIC-simulation results show the process of isochoric heating and thermalization. Subsequently,  $T_e$ declines because of adiabatic plasma expansion. The HD-RT fit, however, shows adiabatic plasma expansion only, which artificially starts with a heuristic initial electron temperature at \SI{0}{\ps} delay. Both approaches are comparable only after thermalization, i.e., at delays later than \SI{1.5}{\ps}. Although the trend of adiabatic cooling by plasma expansion is present for both approaches, the PIC simulation overestimates the bulk-electron temperature compared to the electron temperature of the HD-RT fit. To demonstrate the feasibility of the testbed to benchmark PIC simulations, the next section discusses the disagreement by presenting systematic scans of PIC simulations.

\section{Discussion}
\label{sec:discussion}
\begin{figure}[htb] 
\centering {
\includegraphics[width=83mm]{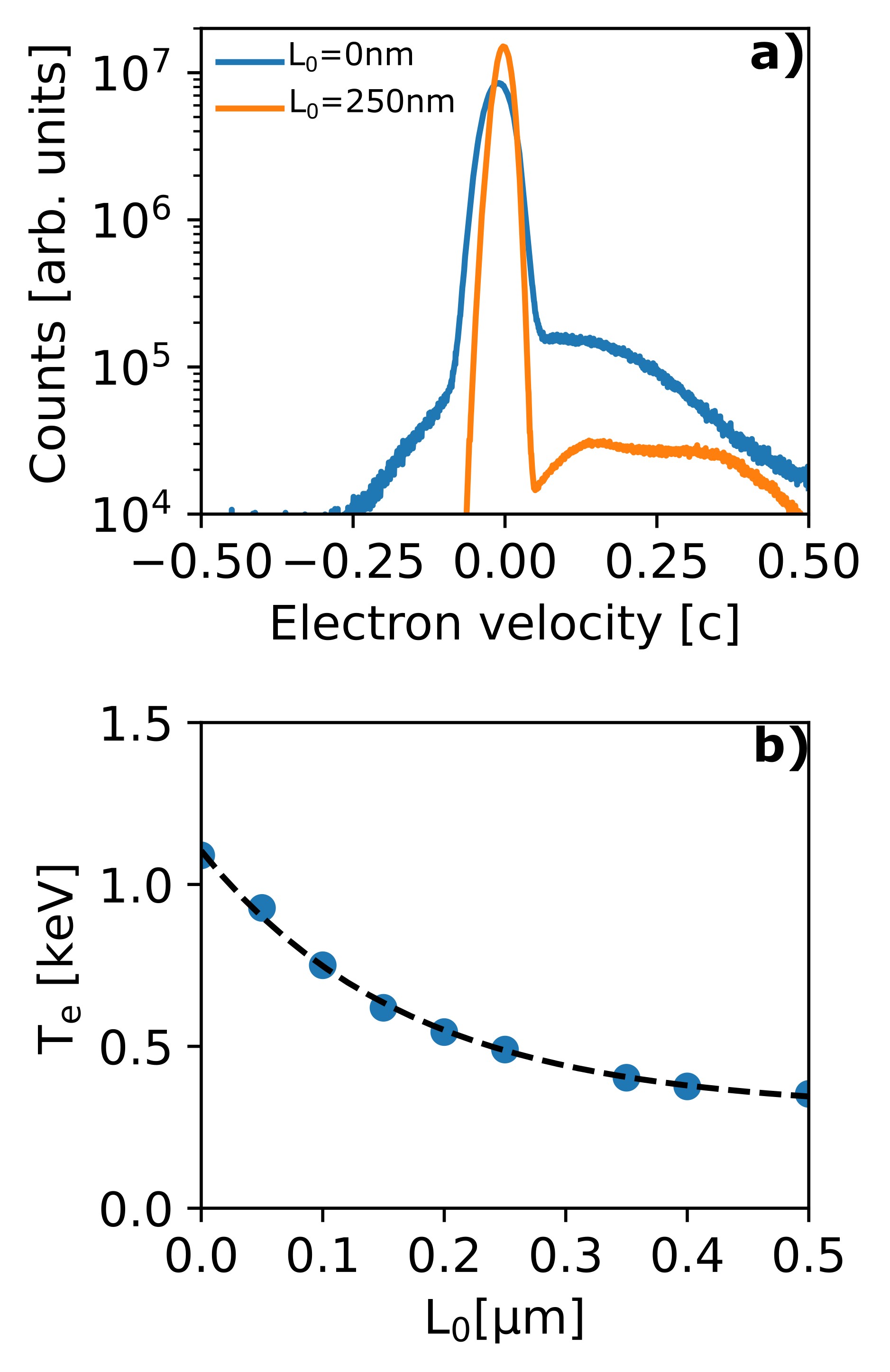}} 
\caption{  PIC simulation results: \textbf{a)}  Electron-velocity distribution in the laser-propagation direction to the time of the maximum return current at \SI{20}{\fs} delay. \textbf{b)} Bulk-electron temperature $T_e$ at \SI{1.8}{ps} delay versus the initial surface scalelength $L_0$. The dashed line is an exponential fit to the data and serve as a guide for the eye.}
\label{Fig.scale_T}
\end{figure}

As exemplified in the previous section, the results of the testbed are feasible to be used for quantitative comparisons to  PIC simulations. In this section we discuss the effect of different parameters of the PIC simulations with respect to the chosen endpoint of the bulk-electron temperature after thermalization. \\
 References \cite{gibbon1992collisionless,brunel1987not,gibbon2005short,kruer2003physics,azamoum2018impact} show that resonance absorption dominates the heating of electrons for laser irradiances $I_{laser} \, \lambda_{laser}^2 \lesssim  \SI{1e18}{W\um^2/cm^2}$ and targets with a surface-density scalelengths $L_0 \gtrsim 0.1 \, \lambda_{laser} = \SI{80}{\nm}$. Vacuum heating of electrons, however,  dominates for targets with a surface scalelengths $L_0 \lesssim 0.1 \, \lambda_{laser} = \SI{80}{\nm}$. Furthermore, Ref. \cite{huang2019maximizing} shows that the existence of preplasma changes the laser-absorption efficiency. The here-utilized peak irradiance is \SI{1.0E18}{W\um^2/cm^2}. Consequently, we expect a dependence of the heating of electron to the surface-density  scalelength of the target.  \\
 According to the experimental uncertainties, PIC simulations with different initial surface scalelengths $L_0$  between \SI{0}{} and \SI{500}{\nm} are conducted. Figure \ref{Fig.scale_T} (b) displays the derived bulk-electron temperatures $T_e$ at \SI{1.8}{\ps} delay. The two limiting cases are $T_e = \SI{1087}{\eV}$ for $L_0 = 0$ and $T_e = \SI{353}{\eV}$ for $L_0 = \SI{500}{\nm}$, which corresponds to a variation of almost a factor of 3.
 An exponential fit to the data (dashed line) suggests the saturation of the temperature decrease with increasing initial scalelength. \\
As explained in the following, the reduction of the bulk-electron temperature for increased surface scalelengths $L_0$ is caused by a change of the laser-absorption mechanism from vacuum heating at small $L_0$ to resonance absorption at higher $L_0$. Figure \ref{Fig.scale_T} (a) shows the velocity distribution of electrons in laser-propagation direction for the cases of $L_0 = 0$ and $L_0 = \SI{250}{\nm}$ at \SI{20}{\fs} delay. Both distributions feature a forward-moving electron current with velocities between $\sim \SI{0.1}{c}$ and \SI{0.5}{c}. However, in the case of $L_0 = \SI{250}{\nm}$, the overall number of forward-moving electrons is  decreased, which is a signature of the transition from vacuum heating to resonance absorption. A second signature of the transition to resonance absorption at higher $L_0$ is given by an  increased temperature of the coronal plasma (red-shaded area in fig. \ref{Fig.3} (b)) for higher $L_0$. The total temporal peak of the forward-moving electron current decreases from \SI{32.4}{kA/\um^2} at $L_0 = 0$ to \SI{10}{kA/\um^2} at $L_0 = \SI{500}{\nm}$. A temporally resolved analytic calculation of the electron-temperature increase by resistive return-current heating based on the simulated forward-moving electron current demonstrates good agreement to the increase of the bulk-electron temperature of the PIC simulations up to \SI{70}{\fs} delay. It follows that the decrease of the forward-moving electron current contributes to a reduction of the bulk-electron temperature as the laser-absorption mechanism changes from vacuum heating at low $L_0$ to resonance absorption at high $L_0$.  \\

The Appendix of this work summarizes  other parameter scans and assumptions of the PIC simulations that influence the bulk-electron temperature, each on a sub-\SI{20}{\percent} level. 
Appendices \ref{app:ionization_state} and \ref{app:radiative_cooling} show that the assumption of an initially fully ionized target and the negligence of radiative cooling are reasonable approximations with insignificant effect on the final evolution of the bulk-electron temperature. 
Appendix \ref{app:2dvs3dPIC} investigates the difference of 2D3V-PIC and 3D-PIC simulations with a high number of particles per cell. The results reveal that, at \SI{0.4}{\ps} delay, the bulk-electron temperature of the 3D-PIC simulation is \SI{14}{\percent} lower than the bulk-electron temperature of the 2D3V-PIC simulation. To furthermore check the influence of the pressure gradient of hot electrons along the $z$ axis in the experimental parameter range, Appendix \ref{app:transverse_gradients} presents 3D-PIC simulations with a larger box size and lower resolution. The results demonstrate that the influence of three-dimensional effects of the hot-electron density distribution are negligible within the FWHM of the pump-laser focal spot. \\
PIConGPU features a Coulomb-binary-collision model, which is given by the Spitzer-resistivity model of return-current Joule heating with a constant cutoff of the collision frequency at low temperatures. However, as shown in Ref. \cite{perez2012improved}, even if a constant cutoff is applied in the low-temperature regime, the Spitzer resistivity strongly deviates from the Lee-More, Redmer, and Monte-Carlo results. Therefore, a low-temperature collision-frequency correction is needed for electron temperatures lower than the Fermi temperature \cite{eidmann2000hydrodynamic}. A calculation of the relevance for the here discussed parameter range is given in Appendix \ref{app:low_temp_coll_correction}.  During the leading edge of the laser pulse, at bulk-electron temperatures below \SI{100}{\eV}, slight deviations between the calculation and the results of the PIConGPU simulation are observed. However, as the collision rate remains unaffected at the hundreds-of-eV temperature level, the low-temperature collision correction has negligible influence on the bulk-electron temperature after thermalization. 
To furthermore test the influence of the analytic uncertainty of the Coulomb logarithm $\ln{\Lambda}$ of $O(\ln{\Lambda}^{-1})$, PIC simulations with a fixed Coulomb logarithm  are performed in Appendix \ref{app:coulomb_log_effects}. A decrease of the Coulomb logarithm  from  \SI{15}{} to \SI{3.5}{} decreases the bulk-electron temperature at \SI{1}{\ps} delay by about \SI{20}{\percent}, which is small compared to the dependence on  the inital surface-density scalelength. \\

In summary, systematic scans of PIC simulations show that mainly the dependence on the initial surface-density scalelength contributes to the overestimation of the bulk-electron temperature by the PIC simulation in section \ref{sec:PICsimulation}. The observed transition over different heating mechanisms of electrons confirms previous work at similar laser-intensity levels \cite{gibbon1992collisionless,brunel1987not,gibbon2005short,kruer2003physics,azamoum2018impact}. Furthermore, the discussion underlines the importance of the exact initial distribution of the target density in PIC simulations that try to model experimental scenarios, which is known, for example, from laser-driven proton acceleration  \cite{Schollmeier2015,Keppler2022}. 

\section{Future prospects}
\label{sec:Outlook}
The here-presented showcase of the testbed utilizes isochoric heating of solid hydrogen by an ultrashort laser pulse with a dimensionless vectorpotential $a_0 \approx 1$. A simple reduction of the pump-laser energy directly leads to a reduction of $a_0$. With that, the showcase demonstrates the readiness of the testbed for controlled parameter scans in experiment and simulation at all laser intensities of $a_0 \lesssim 1$ and varied laser-pulse duration. \\
Furthermore, the testbed is able to investigate the transition from the thermal-driven regime of plasma expansion ($a_0 < 1$) to the hot-electron sheath-driven regime of plasma expansion ($a_0 > 1$) by increasing the pump-laser energy. We expect similarities to the plasma physics of laser-heated nanoparticles that show aspects of hydrodynamic expansion together with Coulomb-explosion \cite{ Varin2012,Gorkhover2016,Nishiyama2019,Niozu2021,Peltz2022}.  At intensities approaching $10^{22} \, \SI{}{\W / \cm^2}$ ($a_0 \gg 1$) similar investigations like in the presented showcase suggest that relativistically induced transparency becomes relevant \cite{Bernert2022}. \\
In the present implementation, the testbed features probe beams at \SI{258}{\nm} and \SI{515}{\nm} wavelength. The probe beams are sensitive to plasma-density gradients around an electron density of about $0.1 \, n_c^{\SI{258}{\nm}} \approx 1 \, n_c^{\SI{800}{\nm}}$ and $0.1 \, n_c^{\SI{515}{\nm}} \approx 0.2 \, n_c^{\SI{800}{\nm}}$ \cite{Bernert2022}. A comparison of both densities to the density profiles of the PIC simulation in figure \ref{Fig.3} (a) shows that the  shadowgraphy diagnostic does not image  the density of the hot coronal plasma, which features electron densities between $0.1 \, n_c^{\SI{800}{\nm}}$ and $\SI{E-3}{} \, n_c^{\SI{800}{\nm}}$ (decreasing with delay). A modification of the experimental setup to probing wavelengths in the near-infrared spectral range (e.g. \SI{2.5}{\um} wavelength) would allow to measure the respective densities and, by this, enable a complementary benchmark of PIC simulations with respect to non-thermal effects beyond the bulk-electron population. \\ 
Cryogenic jets of different material and composition are readily available and frequently used \cite{Kim2018}. In the future, the effect of multi-species mixtures of hydrogen and deuterium will be studied \cite{Huebl2020} and cryogenic Argon-jets will allow to benchmark ionization and recombination dynamics as well as plasma opacitiy, e.g., by probing with extreme-ultraviolet backlighters \cite{ Roedel2012,Wheeler2012, Dollar2013}. \\
Finally, we would like to emphasize that the testbed is ready to be used in combination with other laser-driven secondary sources that induce isochoric heating, for example laser-accelerated ion beams \cite{bang2015uniform}.

\section{Summary and Conclusion}
\label{sec:summary}
In summary, we introduce a testbed to experimentally benchmark PIC simulations based on laser-irradiated micron-sized cryogenic hydrogen-jet targets.  Time-resolved  optical shadowgraphy by two spectrally seperated laser beams measures the temporal evolution of the plasma density. A fitting approach by hydrodynamics and ray-tracing simulations enables the determination of  the bulk-electron temperature evolution after the laser energy was absorbed by the target (\textit{HD-RT fit}). A showcase of the testbed studies isochoric heating of solid hydrogen by laser pulses of \SI{37}{\fs} duration and a dimensionless vectorpotential $a_0 \approx 1$. The HD-RT fit yields a bulk-electron temperature between \SI{250}{} and \SI{300}{\eV}  after absorption of the laser energy, which is supported by systematic scans of PIC simulations. The results  confirm that, due to the interplay of vacuum heating and resonance heating of electrons, an exact determination of the surface-density gradient of the target is key to achieve quantitative agreement of experiments of high-intensity laser-solid interactions with PIC simulations in the regime of $a_0 \approx 1$. The showcase demonstrates the readiness of the  testbed  for controlled parameter scans at all laser intensities of $a_0 \lesssim 1$, which is particularly relevant as PIC-simulation tools develop towards the inclusion of physics models at  subrelativistic laser-intensity levels. By extending the platform with additional multi-color probes and by including diverse atomic target species in the future, the platform establishes a path towards a sophisticated, versatile testbed to systematically explore plasma opacity as well as ionization and recombination dynamics in laser-heated plasmas.

\begin{acknowledgments}
We thank the DRACO laser team for excellent laser support and for providing measurements of the laser contrast. The work of S.A., C.B., I.G., T.K., M.R., U.S., and K.Z. is partially supported by H2020 Laserlab Europe V (PRISES, Contract No. 871124). FLASH  was developed in part by the DOE NNSA- and DOE Office of Science-supported Flash Center for Computational Science at the University of Chicago and the University of Rochester.\\

Author contribution statement: L.Y., L.H., I.G., T.K., X.P., J.V. conducted the simulations. S.A., C.B., S.G., M.R., T.Z., K.Z. conducted the experiments. L.Y. and C.B. wrote the publication and S.A. contributed to section \ref{sec:experiment}, Appendix \ref{app:exp_details} and Appendix \ref{app:initial_diameter}.   U.S. and T.C. supervised the project. All authors reviewed the manuscript. 
\end{acknowledgments}

\newpage
\appendix

\section{Experimental details of the optical microscope}
\label{app:exp_details}
For the \SI{258}{nm}-probe imaging, the magnification is $M=77$ and the measured spatial resolution limit is \SI{<500}{nm}. The utilized camera is a PCO.ultraviolet (14bit CCD sensor with $1392 \times 1040$ pixels of \SI{4.65}{\micro m} size), which results in an overall field of view (FoV) of \SI{84}{\um} $\times$ \SI{63}{\um}. The \SI{515}{nm}-probe imaging has a magnification of $M=70$ with a measured spatial resolution limit of \SI{<1}{\um}. Images are recorded with a PCO.edge 4.2 camera (16bit sCMOS sensor with $2048 \times 2048$ pixels of \SI{6.5}{\um} size each), resulting in a FoV of $190 \times 190$ \si{\um^2}.

\section{Measurement of the variation of the initial target diameter}
\label{app:initial_diameter}
The HD-RT fit is sensitive to the initial diameter of the target. Experimentally, the initial target diameter is defined by the aperture of the source that ejects liquefied hydrogen into the vacuum of the target chamber \cite{kim2016development}. Evaporation of the liquid hydrogen causes the jet to rapidly freeze. This reduces the diameter of the frozen solid hydrogen jet. Energy conservation allows to estimate the amount of liquid hydrogen that is required to be evaporated in order to cause the residual material to freeze. Depending on the initial temperature, the evaporated material constitutes up to \SI{25}{\percent} of the initial liquid volume, resulting in a reduction of the diameter of the solid jet by up to \SI{13.3}{\percent}. In this study, a cylindrical source aperture with a nominal diameter of \SI{5}{\um} and $ \pm \SI{1}{\um} $ manufacturing tolerance is used. This results in an expected initial diameter of the solid hydrogen jet of about \SI{4.3 \pm 0.9}{\um}. \\
To measure the mean and the variation of the target diameter, a bright-field-microscopy  image of the unperturbed target is captured by the \SI{258}{nm} probe and shown in the supplementary figure S 2. The diameter of the target is measured at \SI{62}{} positions that are evenly spaced over the full vertical FoV (blue horizontal lines). The mean target diameter is \SI{4.4}{\um} with a standard deviation of \SI{0.2}{\um}. Furthermore, the image shows typical target-geometry fluctuations.  The target diameter varies  along the jet axis and in the upper part of the image the target is bent to the right side. At  $y=\SI{15}{\um}$ and between $y=\SI{-10}{\um}$ and $y=\SI{-20}{\um}$, the target features structural differences of the bulk compared to  $y=\SI{0}{\um}$, where the target is fully transparent.

\section{Discussion of the HD-RT fit}
\label{app:discussion_HD_RT_fit}

There are two fundamental assumptions of the HD-RT fit. The first one is a homogeneous initial temperature of the bulk-electrons as a result of isochoric heating. Details about this assumption are discussed in section \ref{sec:PICsimulation}. The second assumption is the utilization of hydrodynamics simulations to calculate the plasma-expansion process, which is discussed in this section. We first calculate the Knudsen number $Kn$. Taking the temperature $T_e$ = \SI{300}{eV}, the electron density $n_e$ = $\SI{30}{n_c}^{800nm}$, and the Coulomb logarithm $\ln{\Lambda}$ = \SI{3.55}, the electron-electron-collision rate (for thermalized temperatures $T_e$) calculates to \begin{equation} 
\tilde{\nu}_{ee}=\SI{2.91E-6}{} \, n_e \, \ln{\Lambda} \, T_e^{-3/2} = \SI{1.04E14}{\s^{-1}} \,.
\end{equation}
The thermal velocity of electrons is 
\begin{equation}
v_{the}=\sqrt{\frac{k_B \,T_e}{m_e}} = \SI{7.26E6}{\m / \s} \,,
\end{equation}
with Boltzmann's constant $k_B$ and the electron rest mass $m_e$. The mean free path length $f$ of the electrons is 
\begin{equation}
f= \frac{v_{the}}{\nu_{ee}} = \SI{69.8}{\nm} \,.
\end{equation} 
With the diameter of the target $L = \SI{4.4}{\um} $ as a characteristic spatial scale of the system, the Knudsen number of the bulk plasma is 
\begin{equation}
Kn = \frac{f}{L} \approx \SI{0.02}{} \,,
\end{equation}
which is in the applicable range of hydrodynamics equations \cite{karniadakis2006microflows}. \\
The Debye length $\lambda_D = \SI{1}{\nm}$ gives the 
plasma parameter 
\begin{equation}
\Lambda_\text{plasma} = 4\pi \, n_e \, \lambda_D^3 \gg 1 \, ,
\end{equation}
which shows that the plasma is weakly coupled. The PIC  simulation in section \ref{sec:PICsimulation} yields magnetic fields  $B$ between \SI{1}{T} to \SI{100}{T} in the single-picosecond timeframe and the hydrodynamics simulations yield ion temperatures between \SI{10}{\eV} and \SI{100}{\eV} in the tens-of-picosecond timeframe. The characteristic Lamor radius 
\begin{equation}
r_{L}= \frac{m_i \, v_{thi}}{e \, B\, L} \gg 1 
\end{equation}
shows that the investigated plasma is weakly magnetized during all times. $e$ is the elementary charge. We conclude that hydrodynamics simulations are feasible to calculate the plasma-expansion process. \\

The simulation tool FLASH is commonly used to model high-energy-density physics \cite{waagan2011robust,tzeferacos2018laboratory}. Uncertainties mainly arise from the hydrogen equation of state and the assumption of two-dimensional radial symmetry. Radial symmetry is supported by the experimental observation of similar expansion in a secondary optical-probing axis antiparallel to the pump-laser axis. Lateral heat diffusion by transverse temperature gradients along the $z$ axis is, however, not considered by two-dimensional radial symmetry.  Two-dimensional cylinder-symmetric simulations in Appendix \ref{app:transverse_gradients} shows that the influence is negligible in the investigated parameter range. Different equations of state are considered in Appendix \ref{app:hydrogen_equation_of_state}. \\
As we did not use any laser-plasma-interaction model for the HD-RT fit, the approach  promises to be  robust and versatile. The method does not depend on  specific laser and target parameters and can be readily applied to other laser-target systems. 

\section{Equation of state of hydrogen}
\label{app:hydrogen_equation_of_state}

\begin{figure*}[htb] 
\centering {
\includegraphics[width=0.8\textwidth]{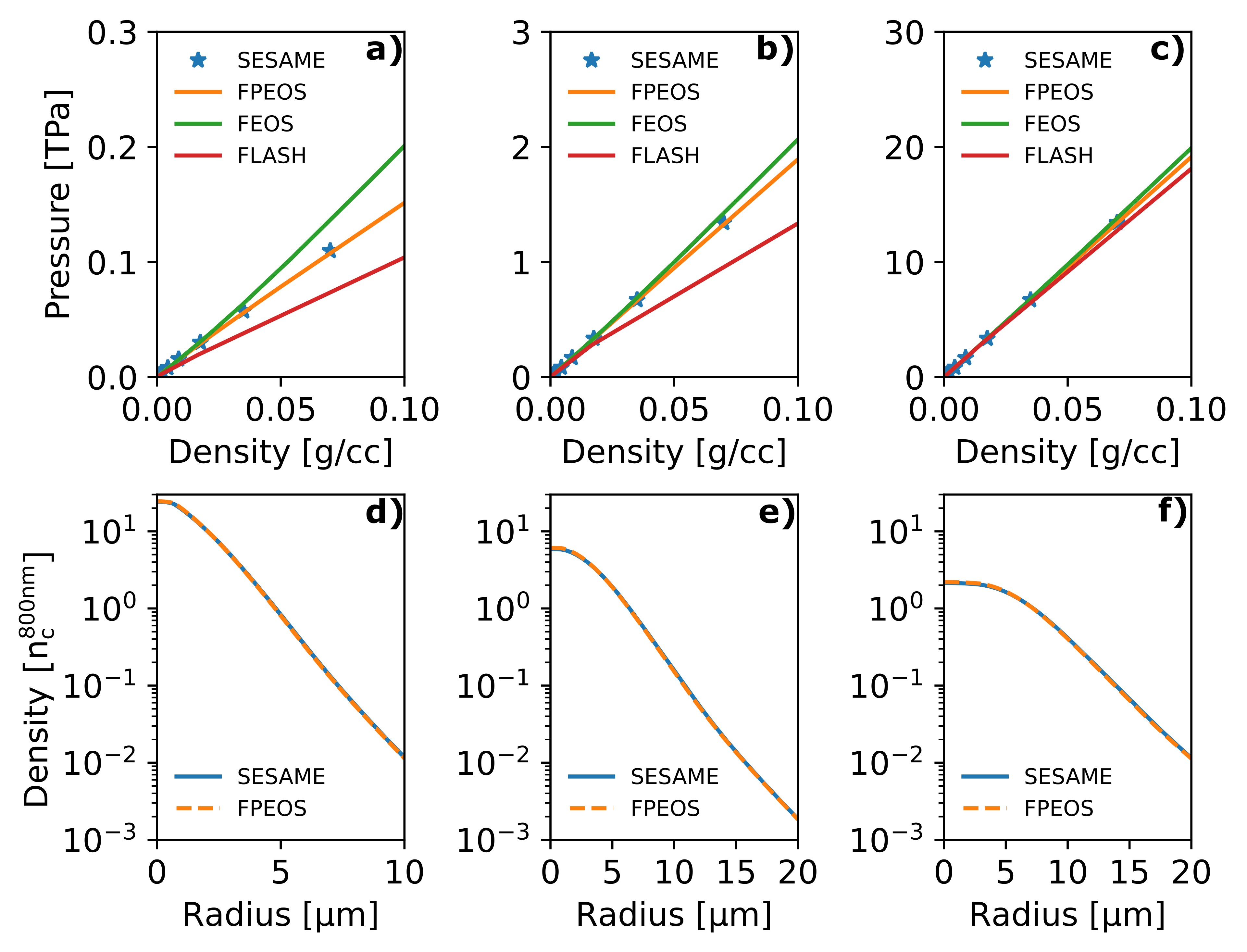}} 
\caption{Comparison of the isotherms of SESAME EOS, FPEOS, FEOS and FLASH EOS of hydrogen for an electron temperature of \textbf{a)}  \SI{10}{\eV}, \textbf{b)} \SI{100}{\eV}, and \textbf{c)} \SI{1000}{\eV}. Comparison of the electron-density profile at \textbf{d)} \SI{10}{\ps}, \textbf{e)} \SI{20}{\ps}, and \textbf{f)} \SI{30}{\ps} as calculated by SESAME EOS and FPEOS. The initial conditions are $T_{e0} = \SI{300}{\eV}$ and $D_0 = \SI{4.4}{\um}$.
} 
\label{Fig.eos} 
\end{figure*}

The equation of state (EOS) is an important input parameter of  hydrodynamics simulations. Here, we test three different hydrogen EOS, which are FPEOS \cite{militzer2021first}, FEOS \cite{faik2018equation}, and FLASH EOS \cite{dubey2009extensible}. SESAME EOS \cite{johnson1994sesame} is used as a benchmark.
We compare the isotherms  of all EOS  at temperatures of \SI{10}{eV}, \SI{100}{eV}, and \SI{1000}{eV}, as shown in the figures \ref{Fig.eos} (a) to (c). The results show that FPEOS fits SESAME EOS best. To compare the plasma-density evolution directly, a FLASH simulation with FPEOS is compared to a FLASH simulation with SESAME EOS in figures \ref{Fig.eos} (d) to (f). The initial settings are $T_{e0} = \SI{300}{\eV}$ and $D_0 = \SI{4.4}{\um}$. The overlapping density profiles confirm the consistency of both EOS. All other FLASH simulations of this work utilize FPEOS.

\section{Ionization state of the target}
\label{app:ionization_state}
In the PIC simulations, a two-dimensional fully ionized plasma column is assumed to resemble the target at the arrival of the laser pulse. According to Ref. \cite{bauer1999exact}, the critical field of barrier-suppression ionization of hydrogen equals a laser intensity of \SI{8e14}{W/\cm^2}. The supplementary figure S 1 shows measurements of the laser contrast via a third-order autocorrelator. No measurement from the same day like the experiment of optical shadowgraphy is available. Both curves show the laser contrast several weeks before and several weeks after the day of the shadowgraphy experiment. Both laser-contrast curves shows that the intensity of \SI{8e14}{W/\cm^2} is already reached at about \SI{400}{\fs} before the laser peak. This reasons an initialization of a fully ionized hydrogen plasma at the starting point of the PIC simulations at  about \SI{-100}{\fs} delay.

\section{Radiative cooling}
\label{app:radiative_cooling}

\begin{figure}[htb] 
\centering {
\includegraphics[width=0.4\textwidth]{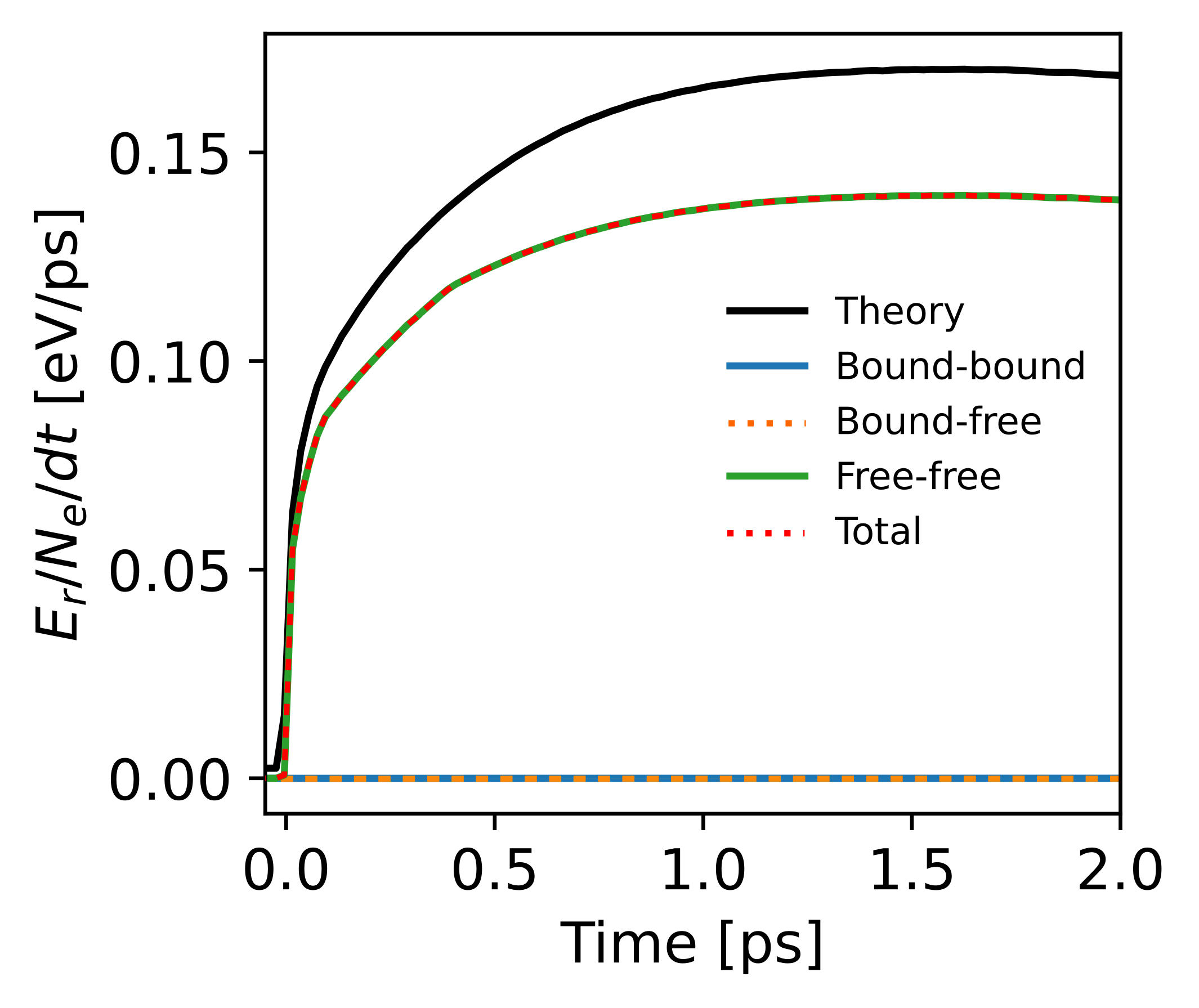}} 
\caption{Energy loss by radiative cooling. The black line corresponds to equation (\ref{eq:radiation_loss}). Calculations by FLYCHK are: bound-bound (blue), bound-free (orange), free-free (green) and total radiation loss (red).  
} 
\label{Fig.FLYCHK} 
\end{figure}

The PIC simulations do not include radiative cooling of the plasma. This Appendix estimates the energy loss by radiative cooling by using analytic calculations and the non-local-thermal-equilibrium tool FLYCHK \cite{chung2005flychk}. We find that the main mechanism of radiation loss is Bremsstrahlung radiation and, as the utilized material is hydrogen with an atomic number of $Z = 1$, the radiation loss per electron is negligible compared to the relevant electron temperatures of several hundred \SI{}{\eV}. \\

To estimate the influence of radiative cooling, we use the results of the PIC simulation (section \ref{sec:PICsimulation}) and calculate the corresponding energy losses from the bulk-electron-temperature evolution. There are mainly three  kinds of radiation.  Bound-bound radiation $P_{Ir}$ accounts for the radiation that is emitted from atomic ionization, free-bound radiation accounts for the radiation by electron recombination 
\begin{equation}
P_{Rr} [W/cm^3] =\SI{1.69E-32}{} \, N_e \, \sqrt{T_e} \, \sum_Z Z^2\, N(Z) \, \frac{E_{\infty}^{Z-1}}{T_e}] \, .                     \label{eq:free-bound}
\end{equation}
$Z$ corresponds to the ionization state. The free-free radiation $P_{Br}$ accounts for Bremsstrahlung radiation by electrons. For a hydrogen-like plasma, $B_{Br}$ is given by 
\begin{equation}
P_{Br} [W/cm^3] = \SI{1.69E-32}{} \, N_e \, \sqrt{T_e} \, \sum_Z Z^2 \, N(Z) \, .                                               \label{eq:bremsstrahlung}
\end{equation}
The total radiation loss is given by the sum of all processes
\begin{equation}
P_r=P_{Ir}+P_{Br}+P_{Rr} \, .                                         \label{eq:Radiation}
\end{equation}
Comparing equations \ref{eq:free-bound} and \ref{eq:bremsstrahlung}, $P_{Rr}$ is $E_{\infty}^{Z-1}/T_e$ times $P_{Br}$. $E_{\infty}^{Z-1}$ is the ionization energy of hydrogen. As the relevant bulk-electron temperatures are in the multi-hundred-\SI{}{\eV} range, $P_{Rr}$ is in the percent range of $P_{Br}$ and is neglected in the following. For fully ionized hydrogen atoms, $P_{Ir}$ is zero. It follows that the total emitted power is approximately equivalent to the power of Bremsstrahlung
\begin{equation}
P_r \approx P_{Br} \, . 
\label{eq:by_Bremsstrahlung}
\end{equation}\\

In the following, we calculate the total radiation loss $E_{r}$ as a function of time $t_0$ from the temperature evolution of the PICLS simulation in figure \ref{Fig.3} (d) by 
\begin{equation}
\begin{split}
E_{r} &=  \int_{0}^{t_0}  P_r(t)\,dt \\
&= \int_{0}^{t_0} \SI{1.69E-32}{} \, N_e \, \sqrt{T_e(t)} \sum_Z Z^2 \, N_i(Z) \,dt \, . 
\label{eq:radiation_loss}
\end{split}
\end{equation}
With $N_i=N_e = 30 \, n_c^{\SI{800}{\nm}} \times \SI{1}{\cm^3}$ and $Z=1$, the total radiation within \SI{2}{\ps} is \SI{1.76e3}{\J}, which is \SI{0.14}{\eV} per electron. Compared to the electron temperature of the PIConGPU simulation at \SI{2}{\ps}, the radiation loss amounts to \SI{0.04}{\percent} only. The temporal evolution is shown by the black line in figure \ref{Fig.FLYCHK}. As radiation loss scales with $Z^2 \, N_e N_i$, the result is in agreement with Ref. \cite{huang2013ion}. \\
The calculation is supported by a calculation with the non-local-thermal-equilibrium tool FLYCHK \cite{chung2005flychk}, for which we use the temperature evolution of the PIC simulation. The result is shown in figure \ref{Fig.FLYCHK}. The bound-bound, bound-free, free-free and total radiation energy are given by the blue, orange, green and red line. Bound-bound and bound-free radiation contribute much less than free-free radiation, which confirms the approximation of equation \ref{eq:by_Bremsstrahlung}. The total radiation loss per electron as calculated by FLYCHK is \SI{0.13}{\eV}. \\
In summary, the main mechanism of radiation loss is Bremsstrahlung (free-free) radiation. Because of the low atomic number of $Z=1$, the radiation loss of an electron is lower than \SI{0.1}{\percent} of the relevant thermal electron energies and radiative cooling can be neglected. 
\color{black}

\section{2D3V-PIC simulation versus 3D-PIC Simulation}
\label{app:2dvs3dPIC}
\begin{figure}[htb] 
\centering {
\includegraphics[width=0.45\textwidth]{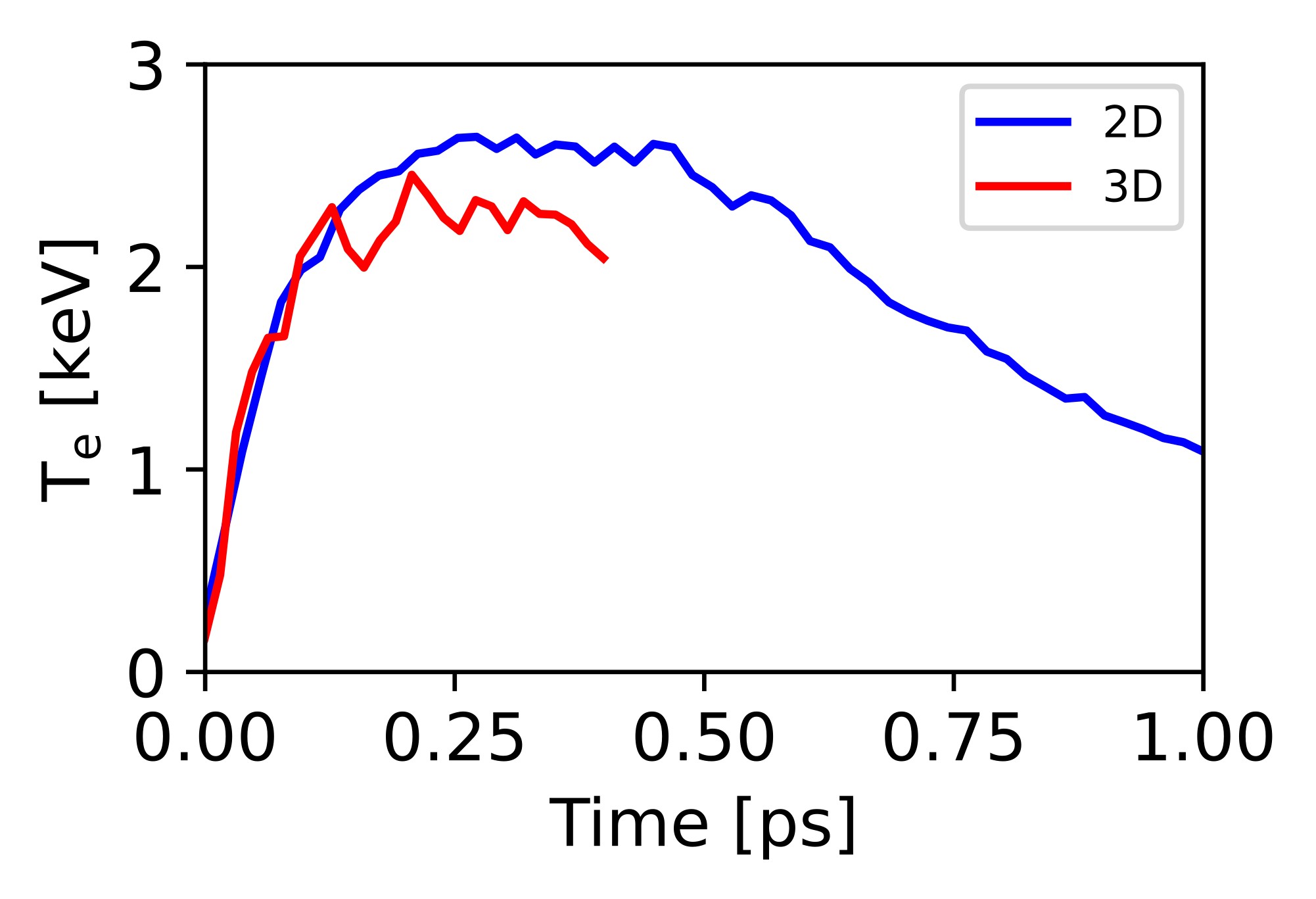}} 
\caption{Comparison of the bulk-electron-temperature evolution of a 3D-PIConGPU simulation and a 2D3V-PIConGPU simulation.} 
\label{Fig.17} 
\end{figure}

To identify possible differences between 2D3V-PIC and 3D-PIC simulations, PIConGPU simulations with a high spatial resolution are compared in the following. To account for limited computing resources, a hydrogen target with \SI{1}{\um} diameter is used. The 2D3V simulation uses a box size of $32 \times \SI{32}{\um^2}$ ($x$, $y$) and the 3D simulation uses a box size of $8 \times 8 \times \SI{12}{\um^3}$ ($x$, $y$, $z$). The total length of the hydrogen target is \SI{4}{\um} into the $z$ direction. To resolve high electron densities in all dimensions, the cell size is  $\SI{0.8}{\um}/96$ and the number of macro particles per cell is \SI{240}{} for both simulations. The 3D simulation is stopped at about \SI{400}{\fs}. \\
For both approaches, a Maxwell-Boltzmann distribution is fitted to the electron-velocity distribution to derive the thermal bulk-electron  temperature. This eliminates the contribution of hot electrons, which is contained in the average kinetic energy. The results from a box size of $5 \times 5 \times 5$ cells in the 3D case and $5 \times 5$ cells in the 2D case are shown in the supplementary figures S 5 and S 6. \\
The comparison of the bulk-electron temperature between the 2D3V-PIConGPU and the 3D-PIConGPU simulation is shown in figure \ref{Fig.17}. Before \SI{100}{\fs}, there is a high agreement between the two simulations. After \SI{100}{\fs}, the temperature in the 2D3V case is increased to slightly higher absolute values. Between \SI{200}{\fs} and \SI{400}{\fs} the difference between the two cases amounts to about \SI{14}{\percent}.

\section{Low-temperature-collision correction}
\label{app:low_temp_coll_correction}

\begin{figure*}[htb] 
\centering {
\includegraphics[width=166mm]{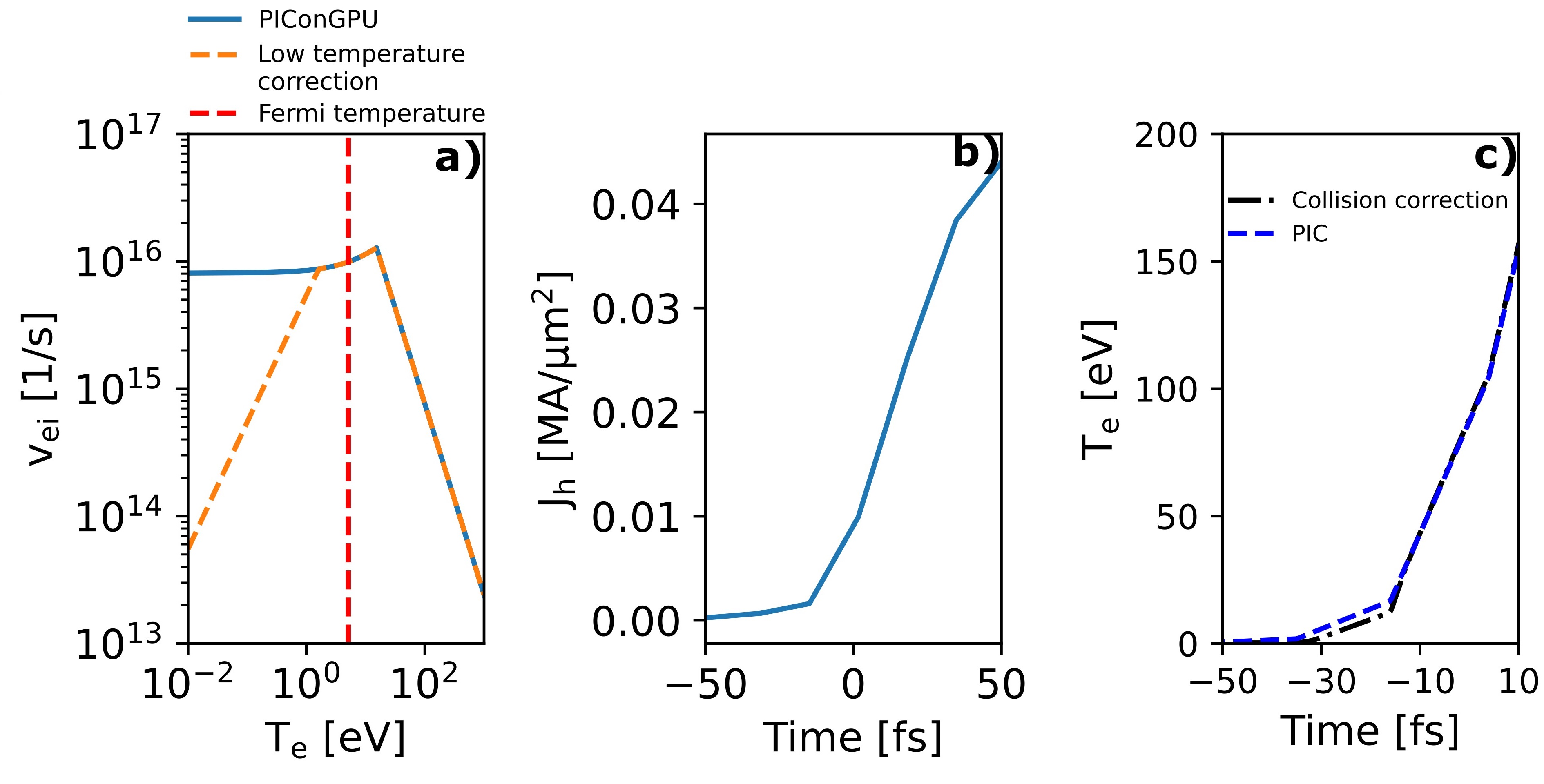}} 
\caption{\textbf{a)} Electron-ion collision frequency versus electron temperature with and without low temperature correction. \textbf{b)} Current density of hot electrons as calculated by equation \ref{eq:joule} from the PIConGPU simulation with a Column logarithm of \SI{5}{} in figure \ref{Fig.13}. \textbf{c)} Electron-temperature evolution from PIConGPU and as calculated from the current density in fig. (b) by taking into account the low-temperature-collision correction.}
\label{Fig.collision_correction}
\end{figure*}

 The collision frequency of the PIC simulations does not include electron-phonon scattering, which is important at electron temperatures below the Fermi energy. However, we can use the temperature evolution as simulated by the PIC simulation, calculate the corresponding current of hot electrons, and reversibly derive the expected temperature evolution by taking the low-temperature-collision correction into account. By comparing the original temperature evolution from the PIC simulation with the artificially calculated temperature evolution, an estimate of the influence of the low-temperature-collision correction can be given. \\
According to the hot-electron scaling in Ref. \cite{wilks1992absorption}, the hot-electron temperature is
\begin{equation}
    T_h=(\sqrt{1+a_0^2/2}-1) \, m_e \, c^2 = \SI{90.5}{\keV}
    \label{eq:ponderomotive_scaling}
\end{equation}
 for $a_0 = 0.88$. Based on Ref. \cite{kemp2006collisional}, the bulk-electron-temperature evolution $\partial T_e / \partial t$ by laser heating is given by 
 \begin{equation}
\frac{3}{2}n_e\frac{\partial T_e}{\partial t}=\frac{\partial}{\partial x}(K_{T_e}\frac{\partial T_e}{\partial x})+\frac{j_h^2}{\sigma_{T_e}}+\frac{3}{2}\frac{n_h \, T_h}{\tau_e} \, .
\label{eq:heating}
\end{equation}
 The three terms on the right-hand side  are diffusion heating ($diff$), resistive return-current heating ($res$), and drag heating ($dra$).   $n_e$ is the electron density, $j_h$ is the current density of hot electrons, $n_h$ is the density of hot electrons, $T_h$ is the average kinetic energy of hot electrons, $\sigma_{T_e}$ is the electron conductivity, $\tau_e$ is the collision time of bulk electrons and $K_{T_e}$ is the thermal conductivity of bulk electrons 
\begin{equation}
  K_{T_e}= \frac{16\sqrt{2}}{\pi^{3/2}} \frac{(k_B \, T_e)^{5/2}}{e^4\, \sqrt{m_e} \, \ln{\Lambda}} \, .
\end{equation}
 Here, $m_e$ is the electron rest mass, $e$ is the elementary charge, $k_B$ is Boltzmann's constant and $\ln{\Lambda}$ is the Coulomb logarithm. \\
  In the PICLS simulation (section \ref{sec:PICsimulation}), a total amount of energy of $\eta_{absorb}=\SI{25}{\percent}$ is absorbed from the laser. The total energy of laser pulse of $E_{laser}=\SI{160}{\mJ}$ and $T_h=\SI{90.5}{\keV}$ gives  the number of hot electrons $N_h$ that is generated by the laser-target interaction
  \begin{equation}
N_h \, T_h = \eta_{absorb} \, E_{laser} \, .
\label{eq:hot_electron}
\end{equation}
  We assume that all the absorbed energy is converted into hot-electron energy. We derive a total number of hot electrons of $N_h\approx \SI{2.8E12}{}$. By assuming a uniform distribution of hot electrons in the plasma column and in the laser spot, we calculate the hot electron density from the corresponding volume $V$ by 
  \begin{equation}
      n_h = \frac{N_h}{V} \approx 7.43\, n_c^{\SI{800}{nm}} = \SI{1.29E22}{\cm^{-3}} \, .
  \end{equation} 
  The current density of hot electrons is 
  \begin{equation}
  j_h = e \, n_h \, v_h \approx \SI{2.48E17}{\A / \m^2}   \, . 
  \end{equation}
    From Ref. \cite{kemp2006collisional} we derive the fraction of temperatures that are generated by resistive versus drag heating $T^{res/dra}$ and resistive heating versus diffusion heating $T^{res/diff}$: 
\begin{equation}
     T^{res/dra} > 22500 \, T_h^{1/3} \, \alpha^{2/3} \, n_{c,23}^{-2/3}  \approx \SI{9E6}{}
\end{equation}
   and 
   \begin{equation}
   T^{res/diff} > 600 \, \alpha^{2/5} \, L_T^{2/5} \approx \SI{3E3}{} \, ,    
   \end{equation}
    with $\alpha=12.9$, $n_{c,23}=0.51$ and $L_T= \SI{4.4}{\um}$.  The equations show that resistive return-current  heating is dominant for the heating of bulk electrons. \\
    Consequently, equation \ref{eq:heating} is simplified to

\begin{equation}
\frac{3}{2}n_c\frac{\partial T_e}{\partial t}=\frac{j_h^2}{\sigma_{T_e}} \, .
\label{eq:joule}
\end{equation}

The electron conductivity is given by
\begin{equation}
\sigma_{T_e}=\frac{n_e \, e^2}{m_e \, \nu_{ei}} \, . 
\label{eq:electron-conductivity}
\end{equation}
The electron-ion collision frequency $\nu_{ei}$ is given by Spitzer's formula 
\begin{equation}
\nu_{ei}=\frac{4}{3} \sqrt{2\pi} \frac{Z_{av}\, e^4 \, m_e \, n_e}{(m_e\, k_B \, T_e)^{3/2}}\ln{\Lambda} \, .
\label{eq:e-i frequency}
\end{equation}
$Z_{av}=1$ in the here considered case of a hydrogen plasma. As the mean free path length of electrons $f$ cannot be smaller than the average ion distance $r_0$
\begin{equation}
 \lambda_e \nless r_0=(\frac{1}{4\pi \, n_i})^{1/3}  
\end{equation}
with the ion density $n_i$, a cut-off at low collision frequencies is introduced 
\begin{equation}
\nu_{cutoff}^{-1} = \frac{r_0}{\sqrt{v_{Th}^2+v_{TF}^2}}  \, .
\label{eq:cutoff}
\end{equation}
$v_{Th}$ and $v_{TF}$ are the electron thermal velocity and the Fermi-temperature velocity. The electron-ion collision frequency including the cutoff correction is 
\begin{equation}
\nu_{ein}^{-1}=\nu_{ei}^{-1}+\nu_{cutoff}^{-1} \, .    \label{eq:correction1}
\end{equation} \\

The low-temperature-collision correction applies to plasma temperatures lower than the Fermi temperature. Here, Spitzer's formula of electron-ion collisions (equation \ref{eq:e-i frequency}) is invalid, because the plasma is in a degenerate state. The collision frequency depends on the scattering of electrons with phonons. The corresponding collision frequency is given by \cite{eidmann2000hydrodynamic}
\begin{equation}
\nu_{ep}=2k_s\frac{e^2 \, k_B \, T_i}{\hbar^2 \, v_{TF}} \, . 
\label{eq:electron-phonon}
\end{equation}
$k_s$ is a constant value that is estimated from experiments, $T_i$ is the ion temperature, $\hbar$ is the reduced Plank's constant. \color{black}  The electron-ion collision frequency $\nu_{eic}$ of a plasma with a temperature smaller than the Fermi temperature is 
\begin{equation}
\nu_{eic}^{-1}=\nu_{ein}^{-1}+\nu_{ep}^{-1} \, .    
\label{eq:correction2}
\end{equation}
The overall electron-ion collision frequency based on equations \ref{eq:e-i frequency} to \ref{eq:correction2} is shown in figure \ref{Fig.collision_correction} (a). For this graph, the Coulomb logarithm is set to \SI{5}{} and the electron density is set to $30 \, n_c^{\SI{800}{nm}}$. The trend of the collision frequency is reversing for temperatures below the Fermi temperature. \\

From equation \ref{eq:joule} we derive a formula of the return-current density  
\begin{equation}
\begin{split}
j_h(t^i) &=\sqrt{\frac{3}{2}n_e \, \sigma_{T_h^i}\frac{\partial T_e^i}{\partial t^i}} \\
& \approx \sqrt{\frac{3}{2}n_c \, \sigma_{T_h^{i+1}}\frac{T_e^{i+1}-T_e^{i}}{\Delta t}} \, , 
\label{eq:current_density}
\end{split}
\end{equation}
with the temporal iteration step $i$ of the following calculation. For each timestep, the electron temperature $T_e$ is derived from the PIC simulation ($\ln{\Lambda}=5$  in figure \ref{Fig.13}). The upper limit of $t$ is set to \SI{50}{\fs}. The retrieved current density of hot electrons is presented in figure \ref{Fig.collision_correction} (b). From the evolution of the hot-electron current $j_h$, the influence of the low-temperature-collision correction on the bulk-electron temperature is calculated and compared to the PIC simulation in figure.\ref{Fig.collision_correction} (c). The artificially derived electron temperature evolution with the temperature evolution of the PIC simulation. Minor differences are observed between \SI{-30}{\fs} and \SI{-10}{\fs}, i.e., before the laser peak arrives on target. After that and due to the rapid Joule heating, both approaches show the same results. It follows  that the low-temperature-collision correction has negligible effect on the final bulk-electron temperature.

\section{Effect of the Coulomb logarithm}
\label{app:coulomb_log_effects}

\begin{figure}[htb] 
\centering {
\includegraphics[width=0.4\textwidth]{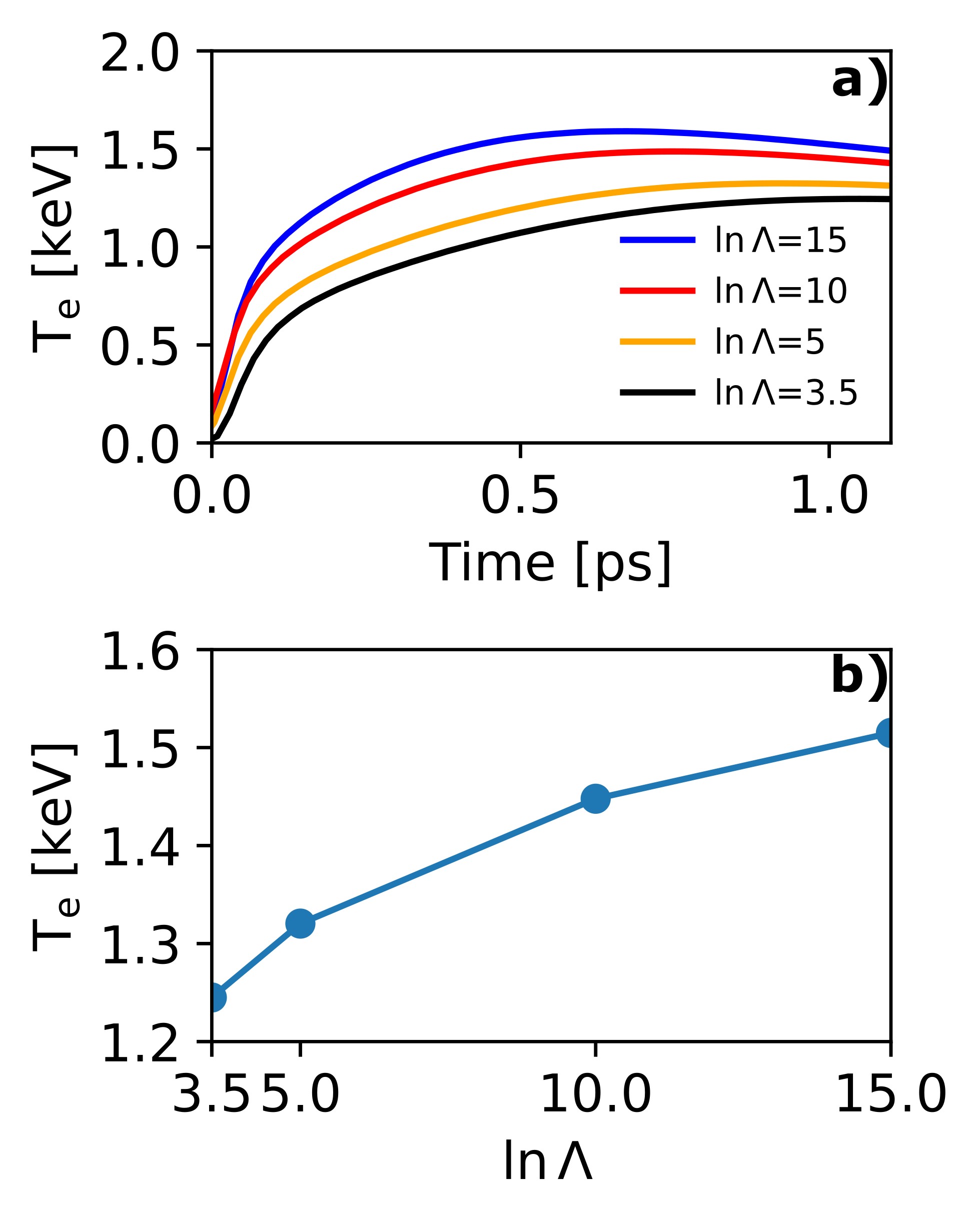}} 
\caption{PIConGPU simulations: \textbf{a)} Bulk-electron temperatures versus time for different fixed values of the Coulomb logarithms $\ln{\Lambda}$. \textbf{b)} Bulk-electron temperature at \SI{1}{\ps} versus $\ln{\Lambda}$}
\label{Fig.13} 
\end{figure}

In the PIC simulations, the energy transfer from laser-heated hot electrons to the bulk electrons is mediated by collisions. The simulations assume a binary collision model, for which the energy transfer is proportional to the Coulomb logarithm $\ln \Lambda$. However, the Coulomb logarithm has uncertainty of $O(\ln{\Lambda}^{-1})$. To test the influence of the Coulomb logarithm, we conduct PIConGPU simulations with different fixed values of $\ln{\Lambda}$ between \SI{3.5}{} and \SI{15}{}. Figure \ref{Fig.13} (a) shows the resulting evolution of the bulk-electron temperature. Figure \ref{Fig.13} (b) compares the derived electron temperatures at \SI{1}{\ps}. A change of the coulomb logarithm from \SI{3.5}{} to \SI{15}{} changes the bulk-electron temperature from \SI{1.25}{\keV} to \SI{1.5}{\keV}, which is a variation of \SI{20}{\percent}.

\section{Lateral heat transfer}
\label{app:transverse_gradients}
\subsection{Hot-electron-pressure gradient in transverse direction}

\begin{figure}[htb] 
\centering {
\includegraphics[width=0.4\textwidth]{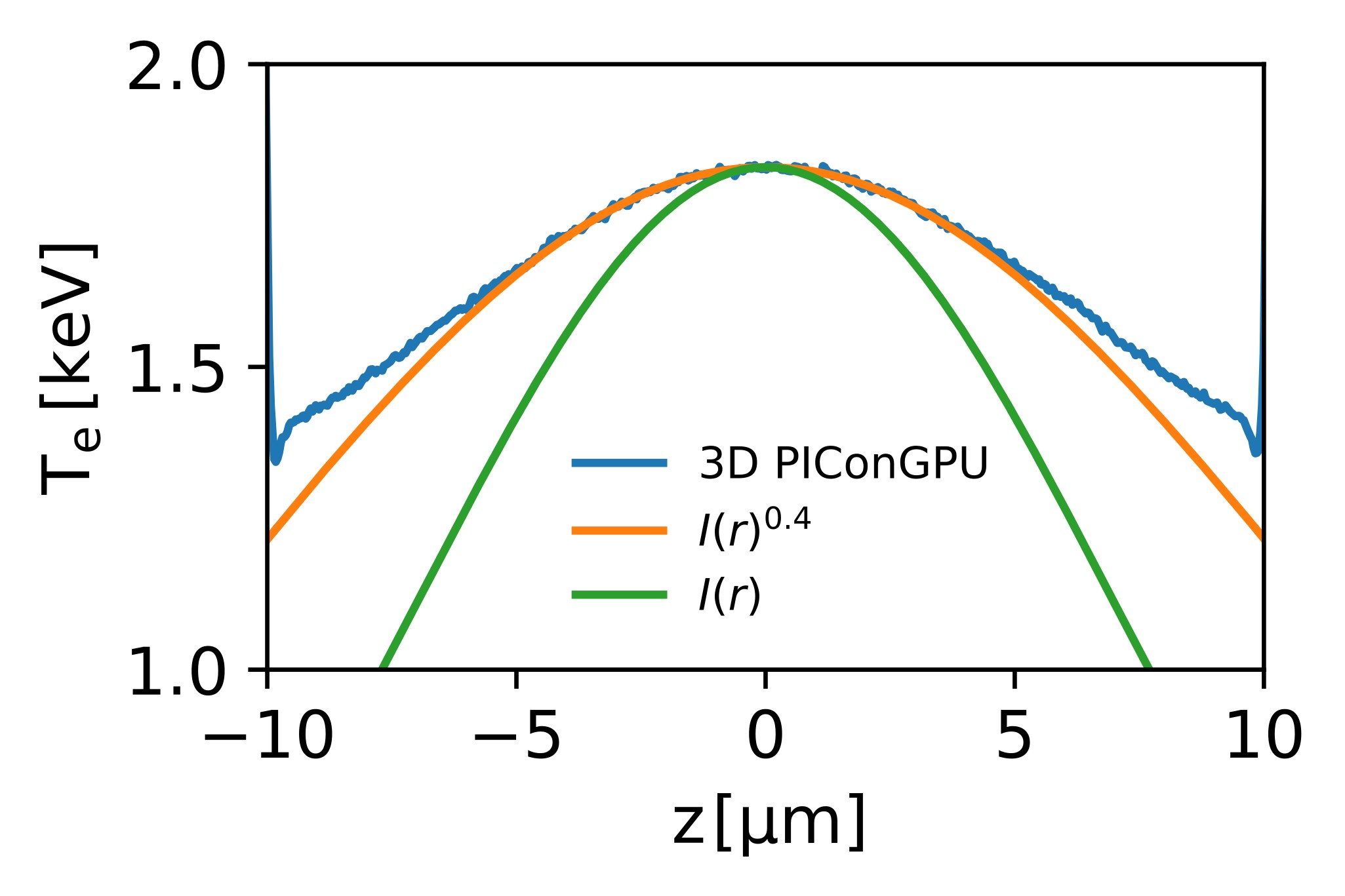}} 
\caption{ Average bulk-electron-temperature distribution along the z axis (jet axis) in the 3D-PIConGPU simulation at \SI{426}{\fs} (blue line). The green and the orange line are the laser-intensity distribution $I(r)$ and $I(r)^{0.4}$, each scaled to the maximum of the blue line.}
\label{Fig.3d_T_gradient} 
\end{figure}

The presented 2D3V-PIC simulations simulate the temperature evolution in the x-y plane only. They ignore the dynamics caused by the hot-electron-pressure gradient in the $z$ direction (along the jet axis). Full 3D-PIConGPU simulations are conducted to investigate the effect of the density gradient of hot electrons to the bulk-electron- temperature distribution within the laser-spot region. The size of the simulation box is \SI{28}{\um} in z direction and \SI{20}{\um} in x and y direction. The initialized target is a cylindrical hydrogen rod of \SI{20}{\um} total length (z direction) and a radius of \SI{2.2}{\um}. The grid size is $\SI{0.8}{\um}/24$ in all directions and the number of macro particles per cell is \SI{240}{}. The initialized hydrogen plasma is fully ionized and features a density of $30 \, n_c^{\SI{800}{\nm}}$ without surface scalelength. The laser parameters are equivalent to section \ref{sec:PICsimulation}. \\
The supplementary figure S 3 shows sectional planes through the center of the target at \SI{426}{\fs} after the laser peak. The bulk electrons are already thermalized at this time. Temperature gradients are observed in all directions. The temperature is decreased from \SI{1.7}{\keV} in the laser-spot region to \SI{1}{\keV} above and below the laser spot (z-axis). Along the laser-propagation direction (y-axis), i.e., into the target bulk, the temperature decreases from \SI{1.7}{\keV} on the front to \SI{1.6}{\keV} on the rear side of the target.  It follows that the temperature distribution in the x-y plane (supplementary figure S 3 (c)) is homogeneous within \SI{20}{\percent}.  This is consistent with the 2D3V-PIC simulation in section \ref{sec:PICsimulation}, for which a uniform temperature distribution is formed after about \SI{500}{\fs}. \\
To estimate the contribution of the hot-electrons-pressure gradient, we calculate the dependency of the local-bulk-electron temperature on the local laser intensity and subsequently compare the calculation to the 3D-PIC-simulation result. In Appendix \ref{app:low_temp_coll_correction} we show that Joule heating by hot electrons is dominant for the heating of bulk electrons. The laser intensity on target is
\begin{equation}
    I(r)=I_0 \, e^{-2\, r^2} \, ,
\end{equation}
with the laser-spot radius $r$. The hot-electron temperature is denoted as $T_h(r)$. Assuming a constant absorption coefficient $\eta$, the density of hot electrons $n_h$ is approximated by equation \ref{eq:hot_electron},  
\begin{equation}
    n_h \, T_h(r) \, dV = \eta \, I(r) \, \tau_0 \, dS \, .
\end{equation} 
Here, $dV$ and $ds$ are the differential volume and area and  $\tau_0$ is the laser-pulse duration. It follows that 
\begin{equation}
    n_h \propto \frac{I(r)}{T_h(r)} \, . 
\end{equation}  
For $v_h \ll c$, the current density of hot electrons is
\begin{equation}
j_h(r)= e \, n_h \, v_h(r) = c \, e \, n_h \, \sqrt{\frac{2 \, k_B \, T_h(r)}{m_e \, c^2}} 
\end{equation}
and 
\begin{equation}
   j_h(r) \propto \frac{I(r)}{\sqrt{T_h(r)}} \, .  
   \label{eq:hot_electron_current}
\end{equation}
 Based on  equations \ref{eq:joule}, \ref{eq:electron-conductivity}, and \ref{eq:e-i frequency} we have 
 \begin{equation}
     T_e(r) \propto \frac{j_h(r)^2}{\sigma_{T_e}} \propto j_h(r)^2 \, T_e(r)^{-\frac{3}{2}}
 \end{equation}
and 
\begin{equation}
    T_e(r) \propto j_h(r)^{0.8} \, .
    \label{eq:hot_electron_diffusion_contribution}
\end{equation}
$T_e(r)$ is the bulk-electron temperature. For $a_0^2/2 \ll 1$, equation \ref{eq:ponderomotive_scaling} gives 
\begin{equation}
    T_h(r) = (\sqrt{1+\frac{a_0^2}{2}}-1) \, m_e \, c^2 \approx \frac{a_0^2}{4} \, m_e \, c^2
\end{equation}
with $a_0 \propto \sqrt{I(r)}$.  We derive 
\begin{equation}
 T_h(r) \propto I(r)   
\end{equation}
and together with equation \ref{eq:hot_electron_current}
\begin{equation}
    j_h(r) \propto \sqrt{I(r)} \, .
\end{equation}
Finally, from equation \ref{eq:hot_electron_diffusion_contribution} we derive the proportionality of the bulk-electron temperature and the intensity distribution of the laser   
\begin{equation}
    T_e(r) \propto I(r)^{0.4} \, .
\end{equation} \\ 
The average bulk-electron temperature variation along the z-axis of the 3D-PIConGPU simulation is shown in figure \ref{Fig.3d_T_gradient} by the blue line. It is calculated from the supplementary figure S 3 (b) by averaging along the x axis. The green and the orange curve in figure \ref{Fig.3d_T_gradient} show  the intensity distribution of the laser $I(r)^{0.4}$ and $I(r)$, each scaled to the maximum of the blue line. Within the FWHM of the laser (\SI{14}{\um}), $T_e$ of  the PIC simulation coincides with the scaling $I(r)^{0.4}$. Beyond this region, the scaling of temperature is slightly different from $I(r)^{0.4}$. This shows that the pressure gradient of hot electrons is relevant only outside the focal-spot FWHM. As the 2D3V-PIC simulations refer to the central x-y plane of the interaction at $z = 0$,  the influence three-dimensional effects of the hot-electron-density distribution can be neglected.

\subsection{Lateral heat transfer by diffusion}
Temperature gradients along the jet axis (z axis) can influence the hydrodynamic expansion of the plasma in the central x-y plane at $z=0$ by lateral heat diffusion. To verify the assumption of two-dimensional radial symmetry of the hydrodynamics simulations of our experimental scenario, we conduct three-dimensional hydrodynamics simulations and compare the plasma density on the tens-of-picosecond timescale for different initial temperature gradients. \\
The simulations utilize FLASH and the simulation box is \SI{20}{\um} in height (z-axis) and \SI{20}{\um} in radius. The initial diameter of the plasma column is \SI{4.4}{\um}. Following the previous subsection, three different temperature gradients are initialized and shown in the supplementary figure S 4 (a). The blue line shows a uniform temperature distribution of \SI{300}{\eV}, the green line shows a temperature distribution proportional to the intensity distribution of the laser $I(r)$ with a peak temperature of \SI{300}{\eV} and the orange line shows a temperature distribution proportional to $I(r)^{0.4}$, similar to the result of the 3D-PIC simulation of the previous subsection. \\
The results of the simulations are presented in the supplementary figures (b), (c), and (d) for \SI{10}{\ps}, \SI{20}{\ps}, and \SI{30}{\ps} delay and at the position of the laser peak at $z=0$. For comparison to the hydrodynamics simulations in section \ref{subsec:hydro}, the red-dashed line shows the results of a two-dimensional radial-symmetric  simulation with $T_{e0} = \SI{300}{\eV}$ and $D_0 = \SI{4.4}{\um}$. All simulations show high agreement to each other. Small deviations of the density profiles occur at densities below $0.1 \, n_c^{\SI{800}{\nm}}$ only and amount to \SI{5}{\percent} at maximum for the case of an initial temperature distribution proportional to $I(r)$. \\
The overall agreement of all three-dimensional cylinder-symmetric simulations to the two-dimensional radial-symmetric simulation shows the negligible influence of lateral heat diffusion in our scenario and justifies the utilization of two-dimensional radial-symmetric hydrodynamics simulations by the HD-RT fit.

\bibliography{ref}
\end{document}